\documentclass[fleqn,usenatbib]{mnras}
\usepackage[]{units}
\usepackage[update,prepend]{epstopdf}
\usepackage{graphicx}
\usepackage{amssymb}
\usepackage{amsmath}
\usepackage{threeparttable}
\usepackage[T1]{fontenc}
\usepackage{hyperref}
\usepackage{color}

\mathchardef\mhyphen="2D

\title[Torus skin outflow in a near-Eddington quasar]{Torus skin outflow in a near-Eddington quasar revealed by spectropolarimetry}

\author[Zakamska and Alexandroff]{Nadia L. Zakamska$^1$\thanks{e-mail: zakamska@jhu.edu}, Rachael M. Alexandroff$^{2,1}$
\\
$^1$Department of Physics and Astronomy, Johns Hopkins University, Bloomberg Center, 3400 N. Charles St., Baltimore, MD 21218, USA\\
$^2$Manhattan District Attorney's Office, New York City, NY, USA
}

\begin{document}
\label{firstpage}
\pagerange{\pageref{firstpage}--\pageref{lastpage}}
\maketitle

\begin{abstract}
Even when the direct view toward the active nucleus is obscured, nuclear emission propagating along other directions can scatter off surrounding material, become polarized and reach the observer. Spectropolarimetry can thus be an important tool in investigating the circumnuclear geometry and kinematics of quasars on scales that cannot yet be probed via direct observations. Here we discuss an intriguing class of quasars where the polarization position angle swings by large amounts ($\sim 90\deg$) within an emission line. We investigate a kinematic model in which the scattering dust or electrons are in an axisymmetric outflow. We propagate Stokes parameters in a variety of geometries of emitter, scatterer and observer. We use these models to predict polarization fraction, line profiles and polarization position angles and compare them to observations. We demonstrate that the swinging polarization angle can be a result of the geometry of the outflow and the orientation of the observer. Polarization properties of a near-Eddington extremely red quasar SDSS J1652 can be successfully explained by a model in which the quasar is surrounded by a geometrically thick disk, whose `skin’ is outflowing at $\sim 1000$ km s$^{-1}$ and acts as the scatterer on scales of a few tens of pc. The line of sight to the observer in this source is within or close to the skin of the torus, in agreement with multi-wavelength data. Spectropolarimetric data and models presented here strongly support the thick-disk geometry of circumnuclear material suggested by recent numerical simulations of high-rate accretion flows onto black holes.
\end{abstract}


\section{Introduction}
\label{sec:intro}

Active galactic nuclei (AGN) powered by accreting supermassive black holes present with a wide range of observational phenomenology. One of the first successful classifications of AGN was based on the presence or absence of broad permitted emission lines and blue continua in their optical spectra \citep{khac74}, resulting in type 1 and type 2 designations. Subsequent spectropolarimetry of type 2 AGN revealed the presence of broad lines and blue continua in the polarized spectra \citep{anto85, mill90}. This key observation gave rise to the geometric unification model of AGN \citep{anto93}, which successfully explains many phenomenological differences between AGN by varying the orientation of the observer relative to optically-thick circumnuclear obscuration. Even if the broad-line region of the AGN cannot be directly seen by the observer due to intervening clouds of gas and dust, some of its emission escapes along other directions, scatters off the surrounding material, becomes polarized and reaches the observer who then detects the nuclear spectrum in the reflected polarized light. 

Scattered light observed using spectropolarimetry has now been used to probe the geometry of AGN in a wide range of objects, from nearby classical Seyfert galaxies \citep{tran92, tran95a, youn96} to more powerful quasars at moderate redshifts \citep{will92, hine93, hine95, smit95, smit00, schm02, zaka05} to high-redshift universe \citep{hine99b, dipo11, cape21}. Some of these studies focused on the dichotomy between the broad-line region and the narrow-line region which are separated by a wide range of scales, with the scattering regions directly visible in high-resolution images of the host galaxy \citep{hine99, zaka06, schm07, obie16}. In contrast, spectropolarimetry of broad absorption line quasars and type 1 quasars can constrain geometry on nuclear scales which cannot yet be probed by any other methods \citep{cape21}. 

It is often difficult to interpret polarimetric observations. Even the dominant scattering agent -- electrons vs dust -- is sometimes problematic to pin down. Electron scattering is favored by a largely wavelength independent scattering efficiency and by the high values of polarization seen in some type 2 AGN, which are in some tension with those achievable in dust scattering. On the other hand, for a standard gas-to-dust ratio dust scattering is more efficient than electrons, and observations of kpc-scale scattering regions where dust is unlikely to be destroyed \citep{hine93, zaka06, obie16} suggest that dust scattering dominates. The situation can get even more complex in spectropolarimetric observations of certain emission lines. In addition to dust and electron scattering, resonant scattering may be important for producing line polarization \citep{lee97, dijk08}. Objects with powerful jets can also be highly polarized, both in the radio and in the optical \citep{ange80, impe90, mead90, smit07b}, but this is due to the synchrotron emission mechanism rather than scattering, and we do not discuss these cases further. 

Velocity structure of the polarization fraction and the polarization position angle has been seen both in narrow emission lines in type 2 AGN \citep{anto85, tran95b} and in broad emission and absorption lines in type 1s \citep{smit02b, smit03, lamy04, dipo13}. In type 2 AGN, the suppression of the polarization fraction within the narrow emission lines relative to the continuum likely indicates that the scatterer and the narrow-line emitter are on similar physical scales, so that the polarization is suppressed by geometric cancellation \citep{tran95b, zaka05}. In contrast, in type 1s the polarization fraction can be suppressed or enhanced across the broad lines \citep{smit02b}, meaning that the continuum-emitting region, the broad-line region and the scatterer can have a variety of size hierarchies. Clearly there is a wealth of information about both the emitter and the scatterer in spectropolarimetric data, but the diversity of observational signatures, geometries and scattering mechanisms can make the interpretation of spectropolarimetric observations very complicated. 

In this paper we develop a kinematic model of an axisymmetric scattering region or wind which allows us to model the velocity structure of the scattered and polarized light as seen in optical and ultra-violet (UV) emission lines for comparison with spectropolarimetric observations. A class of objects of particular interest to us is extremely red quasars (ERQs), a fascinating exclusively high-redshift $(z\sim 2)$ population which was identified by their high infrared-to-optical ratios, extremely high bolometric luminosities reaching $10^{48}$ erg s$^{-1}$ and peculiar rest-frame UV spectra with oddly shaped, high equivalent width emission lines \citep{ross15, hama17}. Upon follow-up observations, the ERQs turned out to have the fastest outflows of ionized gas seen in the [OIII]$\lambda$5008\AA\ emission line of any quasar population \citep{zaka16b, perr19}. These outflows are now unambiguously detected on galaxy-wide scales \citep{vayn21a, wyle22, vayn23b, vayn23c} and are therefore extremely powerful and suspected of undergoing the long-sought `blow-out' phase of quasar feedback on the host galaxy \citep{hopk10}. Their activity is not particularly associated with powerful radio jets: while their radio luminosity is in the radio-intermediate regime, the majority are point-like \citep{hwan18}, and their radio emission is consistent with being a byproduct of winds \citep{zaka14}. Both their extremely high luminosities and their extreme outflow activity suggest that they are near- or super-Eddington sources \citep{zaka19}. These objects are also very highly polarized in the rest-frame UV, with peculiar kinematic structure likely reflecting the geometry of the circumnuclear gas flows \citep{alex18}. 

With spectropolarimetric observations and modeling we can hope to resolve the internal kinematics of the emission and scattering regions which may not be accessible via any other techniques. In Section \ref{sec:phen} we introduce the observational phenomenology we are aiming to explain. In Section \ref{sec:model} we present the model setup, in Section \ref{sec:analysis} we discuss model results and comparisons with observations, in Section \ref{sec:disc} we discuss the implications of our results and we conclude in Section \ref{sec:conc}. Emission line wavelengths are given in vacuum. Ground-based observations are converted onto the vacuum wavelength scales using \citet{mort91}. The orientation of polarization is defined by the orientation of the electric field $\vec{E}$ in the scattered electromagnetic wave. Polarization position angles (PAs) $\beta$ are measured East of North, with $Q=1$ and $U=0$ corresponding to the $E$-vector of polarization oriented in the North-South direction. We use lower case $q$ and $u$ for fractional polarization. The scattering angle $\psi$ is defined as the angle between the wave vectors of the incident and the scattered photons. We use a flat $\Omega_m=0.3$, $\Omega_{\Lambda}=0.7$, $h=0.7$ cosmology for computing distances and luminosities.  

\section{Observational motivation}
\label{sec:phen}

\subsection{Extremely red, near-Eddington quasar SDSS~J1652}
\label{sec:1652}

Our prototypical target is SDSS~J165202.64+172852.3, hereafter SDSS~J1652, an ERQ \citep{ross15, hama17} at $z=2.9$ originally selected by its high infrared-to-optical ratio. When classified based on its UV and optical emission-line properties, it is a high-redshift type 1.8$-$2 (obscured) quasar candidate, with a high equivalent width of CIV$\lambda$1550\AA\ and a high [OIII]$\lambda$5008\AA/H$\beta$ ratio \citep{alex18}. The width of its CIV emission line (full width at half maximum$=2400$ km s$^{-1}$) places it just above the standard cutoff ($<2000$ km s$^{-1}$) for type 2 quasar selection \citep{alex13}, and there is a weak broad component in H$\beta$ \citep{alex18} and a stronger broad component in H$\alpha$ \citep{vayn23c}. X-ray observations confirm that the source is highly obscured, with a column density between the observer and the X-ray emitting corona of $N_H\simeq 10^{24}$ cm$^{-2}$ \citep{goul18a, ishi21}. 

As other ERQs, SDSS~J1652 shows a blueshifted and broad (velocity width containing 80\% of line power $w_{80}=1760$ km s$^{-1}$; \citealt{alex18}) [OIII]$\lambda$5008\AA\ emission line indicative of outflow activity on scales $\gg 100$ pc where this line cannot get collisionally de-excited. The object was therefore observed by {\it James Webb Space Telescope} in its first weeks of science operations as part of an Early Release Science program (``Q-3D'', PI: Wylezalek) to investigate quasars with strong galaxy-scale outflows. High-velocity (several hundred km s$^{-1}$) outflows in this source have now been detected over the entire extent of the galaxy \citep{vayn21b, wyle22, vayn23b, vayn23c}. The object inhabits a massive (a few $L_*$) galaxy with multiple companions and tidal tails indicating merging activity \citep{zaka19, wyle22}. It is a point-like optically-thin radio-intermediate source with no evidence for powerful radio jets \citep{hwan18}. 

Spatially resolved kinematic maps from {\it JWST} allow identification of narrow emission lines associated with the ionized gas in the host galaxy, giving a precise redshift of $z=2.9489$ \citep{wyle22}. It is in excellent agreement with that based on the narrow component of [OIII] seen in the spatially integrated ground-based data \citep{alex18}. All velocities are hereafter measured relative to this frame. 

The bolometric luminosity of the source of $5\times 10^{47}$ erg s$^{-1}$ \citep{wyle22} corresponds to the Eddington luminosity of a $4\times 10^9$ M$_{\odot}$ black hole. While there is no independent method for measuring black hole masses for objects with strong outflow activity and unknown and possibly appreciable obscuration levels, this value is at or above the maximal mass of black holes at present epoch \citep{gult09}, so we assume that SDSS~J1652 cannot be much more massive and the Eddington ratio must be close to or exceed unity. 

\begin{figure*}
\includegraphics[width=\textwidth, clip=true, trim=0 0.5cm 0 0.5cm]{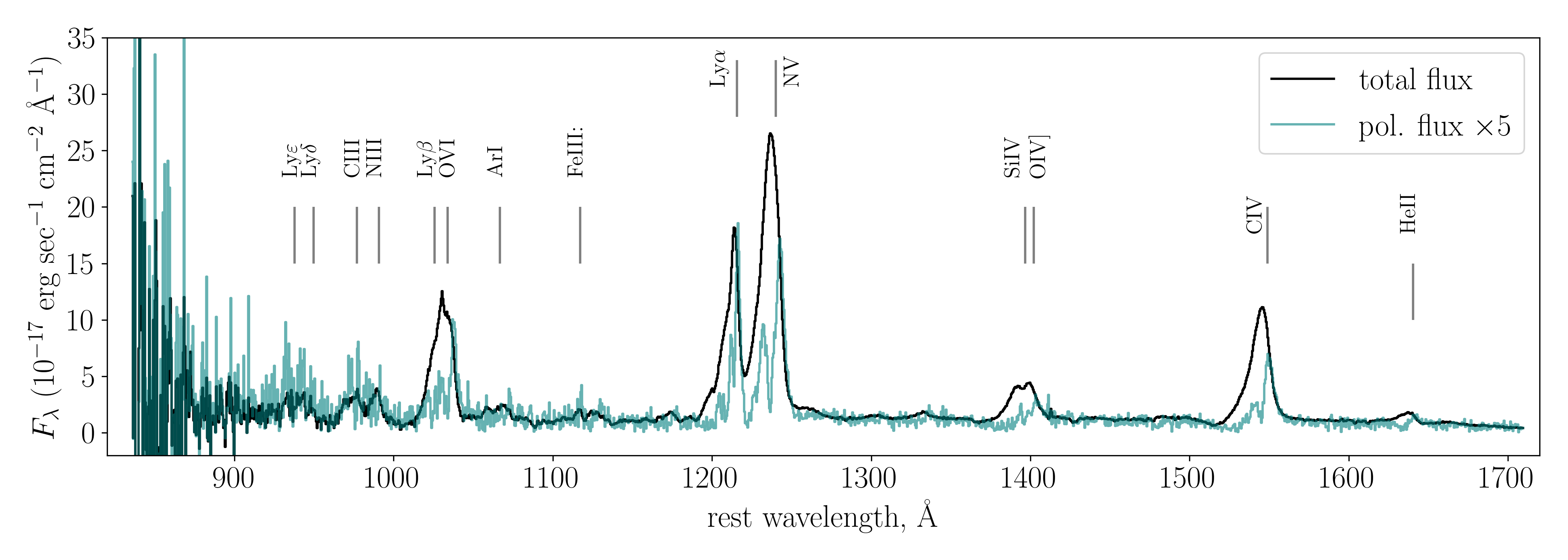}\\
\includegraphics[width=\textwidth, clip=true, trim=0 0.5cm 0 0.5cm]{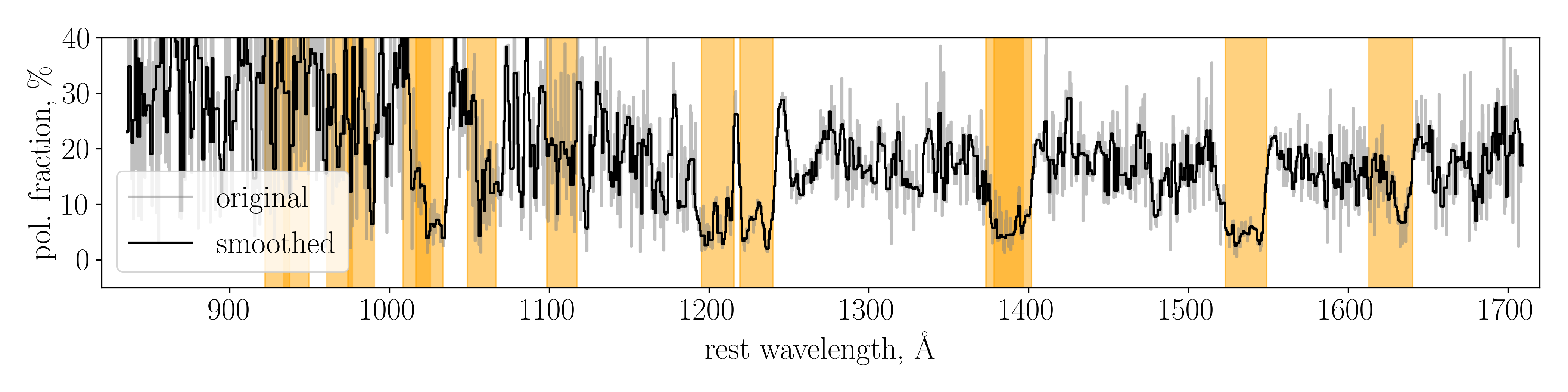}\\
\includegraphics[width=\textwidth, clip=true, trim=0 0.5cm 0 0.5cm]{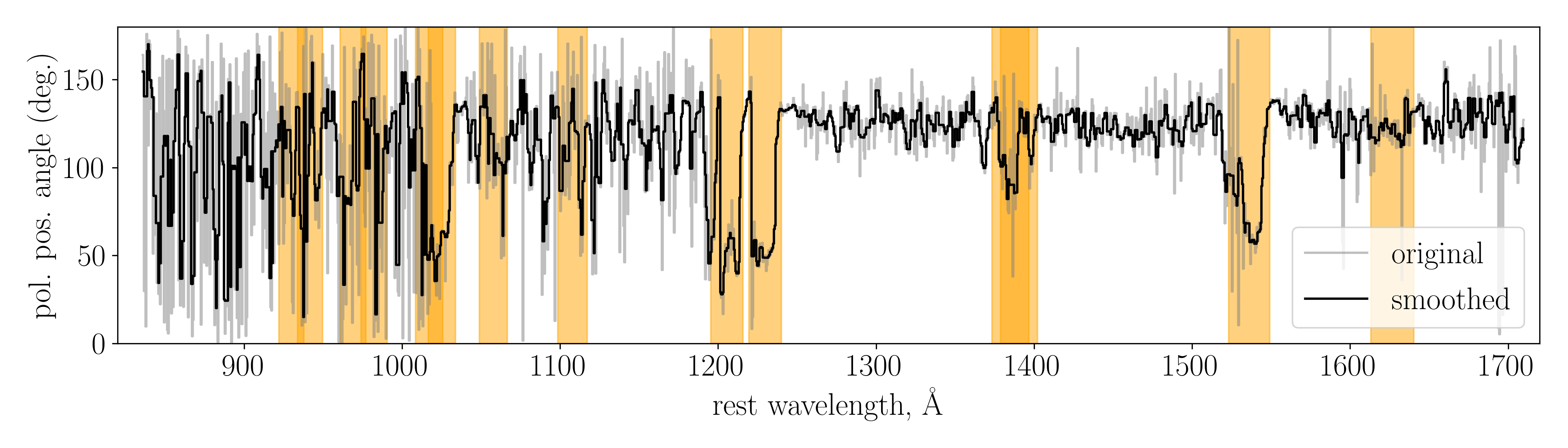}
\caption{{\it Top:} The total (black) and the polarized (teal) Keck LRISp optical spectra of SDSS~J1652. Identified emission lines (based on the line list from \citealt{vand01}) are shown with vertical lines and labels. Their indicated expected rest-frame position is computed based on the redshift of the narrow-line rest-frame optical emission lines observed by {\it JWST} \citep{wyle22}. {\it Middle:} Polarization fraction, both original (grey) and smoothed with a 5-pixel median filter (black). Orange bands show the $(-5000,0)$ km s$^{-1}$ velocity range for each emission line. {\it Bottom:} Polarization position angle, both original (grey) and that obtained from $Q_{\lambda}$ and $U_{\lambda}$ smoothed with a 5-pixel median filter (black). }
\label{pic:1652}
\end{figure*}

\begin{figure*}
\includegraphics[width=\textwidth ]{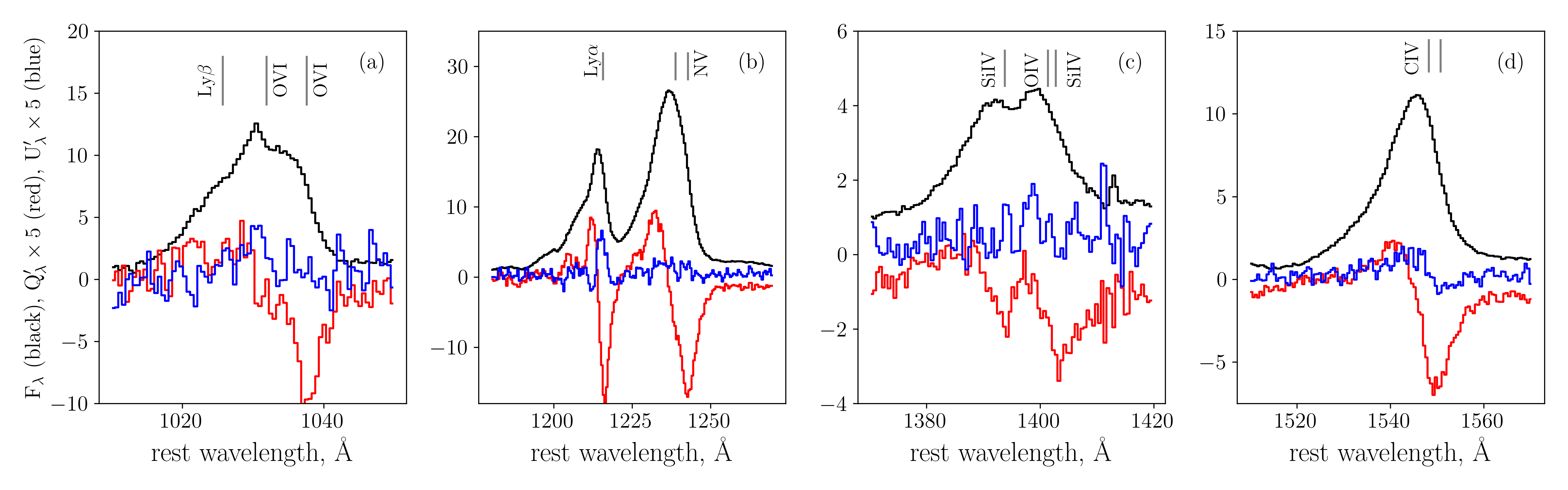}
\caption{The kinematic structure of the Stokes parameters within four major emission line complexes in SDSS~J1652. The coordinate system has been rotated so that $Q'$ (red) and $U'$ (blue) are measured relative to $\beta_0=45\deg$ E of N, the primary axis of illumination based on the large-scale scattered light and photo-ionization observations \citep{vayn21a}.}
\label{pic:1652_lines}
\end{figure*}

In Figures \ref{pic:1652} and \ref{pic:1652_lines} we show spectroscopic and spectropolarimetric observations of SDSS~J1652 obtained using Keck's Low Resolution Imaging Spectrometer. Data acquisition and processing to obtain $F_{\lambda}$, $Q_{\lambda}$ and $U_{\lambda}$ are described by \citet{alex18}. We identify the following important polarization properties:
\begin{enumerate}
\item All emission lines are blueshifted relative to the host galaxy rest frame known from other data; the velocity offsets of the peaks from the expected position range between $-$450 and $-$850 km s$^{-1}$. 
\item While the UV and optical continuum and emission lines are well detected, the high level of continuum polarization ($\sim 20\%$) suggests that the entire UV emission may be due exclusively to scattered light without any direct contributions from the broad-line region and continuum. 
\item The polarization position angle varies dramatically within the emission lines, `swinging' by as much as 90 $\deg$. 
\item As a function of velocity, the pattern is similar within all emission lines: the polarization position angle is the same in the red part of the line as the position angle of the continuum, and the `swing' affects the blue part. 
\item The degree of polarization is smaller in the blue part of the emission line than in the red part.
\item The peak of the polarized intensity is redshifted in comparison to the peak of the total line flux. 
\end{enumerate}

In the rest-frame UV continuum imaging of the object using {\it Hubble Space Telescope}, there is a clear detection of an extended nebula to the Southwest of the nucleus, with a possible fainter counterpart to the Northeast \citep{vayn21b}. This nebula is orthogonal to the continuum polarization position angle (125 $\deg$), and therefore it is interpreted as the scattered light from the region of the host galaxy illuminated along the polar direction of circumnuclear obscuration, by analogy to low-redshift type 2 quasars \citep{zaka05, zaka06}. The gas with the highest ionization levels -- photoionized by direct quasar emission -- is oriented in the same direction, further supporting this interpretation. The Southwestern region is likely tilted toward the observer and appears brighter and detectable in ground-based data \citep{vayn21b}, whereas the Northeastern region is pointed away from the observer and required {\it JWST} data to identify \citep{vayn23c}, likely because it is extincted by the intervening dust. 

Thus the orientation of the polar axis of circumnuclear obscuration is known from the spatially resolved observations. We take this axis to be at position angle $\beta_0=45\deg$ East of North and we apply a rotational transformation to the observed $Q$ and $U$ Stokes parameters to obtain $Q'$ and $U'$ relative to this new axis, which we show in Figure \ref{pic:1652_lines}. The continuum polarization has values $Q'<0$, which means it is orthogonal to the chosen axis of origin, but within emission lines $Q'$ `swings' to positive values on the blue wings. $U'$ values -- which reflect values of polarization at $\pm 45\deg$ to the chosen axis -- are close to zero. 

\subsection{Other objects with swinging polarization angle}

Some of the tell-tale spectropolarimetric properties listed above are manifested by other quasars. Of the four objects other than SDSS~J1652 presented by \citet{alex18}, two more show dramatic changes in the polarization position angle across the emission lines. One (SDSS~J1515+1757) is a classical type 2 quasar both at rest-frame UV and at optical wavelengths as identified by the width of the emission lines \citep{alex13}. The other one (SDSS~J1623+3122) satisfies both type 2 and ERQ selection criteria and shows clear evidence of [OIII]$\lambda$5008\AA\ ionized gas outflow in its infrared spectrum.  Both objects are polarized at the $\sim 10\%$ level. There are interesting differences between the properties of these two sources and SDSS~J1652 -- in particular, in both cases the `swinging' part of the line is redshifted -- by $300-600$ km s$^{-1}$ -- relative to the total line centroid, instead of blueshifted as it is SDSS~J1652. 

Mrk~231 is a nearby reddened FeLoBAL quasar which has an UV absorption system blueshifted by $\sim$4,600 km s$^{-1}$ with respect to the emission lines \citep{smit95}, indicative of a dusty, outflowing BAL screen with a covering factor of 90$\%$ \citep{veil16}. In spectropolarimetric observations of a broad H$\alpha$ emission line, the polarization position angle changes by 10$^{\circ}$ over line in a characteristic `S'-shaped pattern, with the polarization position angle of the redshifted wing of H$\alpha$ roughly matching the continuum polarization position angle. The polarization fraction is about 3.5\% in the continuum, dips to 2.5\% on the blue wing of H$\alpha$ and increases to nearly 5\% on the red wing. As a result, the polarized line intensity is redshifted by about 600 km s$^{-1}$ compared to the overall line intensity \citep{good94, smit95}. 

PG1700+518 is a low-redshift ($z=0.288$) BAL quasar with a set of UV absorption systems at velocities ranging between 7,000 and 18,000 km s$^{-1}$ \citep{pett85}. In optical spectropolarimetry, both the broad H$\alpha$ and to a lesser extent H$\beta$ show features similar to those of SDSS~J1652. Relative to the 1\% continuum polarization, the polarization fraction dips to 0.5\% on the blue wing of the line and increases on the red wing of the line to 1.5\%, resulting in a net redshift of nearly 4,000 km s$^{-1}$ of the polarized line profile relative to the total line intensity \citep{youn07}. The polarization position angle (PA) rotates by $\sim90\deg$ within the emission line, though unlike SDSS~J1652, the PA within the emission line does not match the continuum PA either on the redshifted or on the blueshifted wing of the line profile. 

The presence of broad Balmer emission lines and the relatively low levels of polarization seen in the latter two sources (1$-$2\%) suggest that despite intervening absorption and extinction, a large fraction of the observed optical and UV emission is due to directly seen nuclear continuum and circumnuclear broad-line region. This component is in general unpolarized, so it dilutes the scattered light signal to relatively low polarization values. In obscured quasars of \citet{alex18}, the direct unpolarized light from the nucleus is hidden and the levels of polarization are correspondingly higher ($\sim 10\%$ and above). 

Previously some of these spectropolarimetric features were modeled by a rotating disk wind (e.g., in the models of Seyfert 1 galaxies and PG1700+518 by \citealt{youn00,smit05,youn07}) or by multiple scattering components (e.g., in Mrk~231 by \citealt{smit95}). In light of the extreme polarization properties of SDSS~J1652 (large PA swing, high levels of polarization) we revisit spectropolarimetric modeling in this paper and explore various scattering geometries in search of a natural explanation for the observed spectropolarimetric phenomenology. As some of the objects of interest show strong outflows, we are particularly interested in models that can simultaneously account for the outflow activity and the polarization properties. 

\subsection{Polarization mechanism}
\label{sec:resonant}

In the quasars we discuss here, which are not dominated by jet emission, polarization is expected to be dominated by scattering. Scattering can be produced by free electrons (Thomson scattering), dust \citep{drai03}, or, for emission lines, by partly ionized gas which resonantly scatters incident photons if they arrive with the right range of energies to match the transition in question. There are some qualitative similarities between all three mechanisms -- in particular, the dominant emerging polarization is perpendicular to the scattering plane. This property of scattering allowed for major break-throughs in understanding the geometry of AGN obscuration \citep{anto85}.

But there are also qualitative and quantitative differences. Resonant scattering can have high optical depth per small column density of gas, but does not work for continuum emission. When dust is present, even in fully ionized gas it is a more efficient scatterer than electrons, by up to two orders of magnitude depending on the wavelength, but because of the admixture of particle sizes it is a less efficient polarizer than electrons. All three types of scatterers are known to be important in active nuclei, depending on spatial scales and observed wavelengths: electron scattering may dominate on small scales close to the nucleus where dust is destroyed and / or at wavelengths where dust is an inefficient scatterer \citep{gall97}, dust scattering may dominate in the optical for scattering on galactic scales \citep{zaka06}, and Ly$\alpha$ resonant scattering in the surrounding gas-rich galaxy and halo may have a major effect on the observed emission line profiles \citep{dijk06}. 

In this paper we analyze both line and continuum polarization of a class of quasars defined by certain similarities in their polarization properties, so our first task is to determine which, if any, scattering process is likely to dominate. In \citet{alex18}, we noted the similarity between the overall polarization fraction and polarization position angle in the continuum and in the red wings of the emission lines, which argued in favor of the same polarization mechanism for the lines and the continuum and therefore against resonant scattering. We then found that resonant scattering optical depth could be high, e.g., for CIV$\lambda$1550\AA\ emission $\tau_{\rm res}/\tau_{\rm dust}\simeq 300\times \eta_{\rm CIV}$ for a typical range of gas velocities ($\sim 3000$ km s$^{-1}$). This calculation assumes that gas and dust are well mixed, and $\eta_{\rm CIV}$ is the fraction of carbon in the relevant ionization state, up to 0.3. Because $\eta$ is sensitive to the location and physical conditions of the scatterer, we left the question of the importance of resonant scattering unresolved. 

We tackle the contribution of resonant scattering again, now using observations of multiple different emission lines in SDSS~J1652 as a new constraint (Figure \ref{pic:1652_lines}). For resonant scattering the polarization fraction as a function of scattering angle $\psi$ is
\begin{equation}
p_{\rm res}(\psi)=\frac{p_0\sin^2\psi}{1+p_0\cos^2\psi}.
\end{equation}
Here $p_0$, the maximum level of polarization achieved at $\psi=90\deg$, depends on the angular momentum quantum numbers $J_e$ and $J_g$ of the excited and the ground state and is tabulated by \citet{hami47} and \citet{lee94a}. Crucially, $p_0=0$ for the following three combinations of $J_g$ and $J_e$: (i) $J_e=0, J_g=0$; (ii) $J_e=0, J_g=1$; and (iii) $J_e=1/2, J_g=3/2$. Therefore, if there are polarized transitions with these values in our spectra, we can rule out resonant scattering as the dominant mechanism. 

We use the NIST atomic spectra database\footnote{\url{https://www.nist.gov/pml/atomic-spectra-database}} to record the $J_e$ and $J_g$ values for all transitions shown in Figures \ref{pic:1652} and \ref{pic:1652_lines}. Several features (Ly$\beta$, Ly$\alpha$, NV and CIV) are mixes of $J_e=1/2\rightarrow J_g=1/2$ ($p_0=0$) and $J_e=3/2\rightarrow J_g=1/2$ ($p_0=0.429$) transitions with equal or similar wavelengths and Einstein coefficients, so the resulting scattering can be polarized and therefore they do not provide a clean test of the resonant scattering mechanism. 

The `smoking gun' feature turns out to be the blend of SiIV 1393.8\AA\ ($J_e=3/2\rightarrow J_g=1/2$ with $p_0=0.429$), SiIV$\lambda$1402.8\AA\ ($J_e=1/2\rightarrow J_g=1/2$ with $p_0=0$) and OIV$\lambda$1401.4\AA\ with $J_e=1/2\rightarrow J_g=3/2$ ($p_0=0$), $J_e=1/2\rightarrow J_g=1/2$ ($p_0=0$) and $J_e=3/2\rightarrow J_g=5/2$ ($p_0=0.015$). In Figure \ref{pic:1652_lines}(c), the line is clearly highly polarized, with the redder component (made up of SiIV$\lambda$1402.8 and OIV$\lambda$1401.4) showing polarization levels (close to 20\%) similar to those or higher than those of the bluer component. In contrast, if resonant scattering were the dominant polarization mechanism, the red part would be expected to be unpolarized or polarized at a very low level, since the only potentially polarized scattering would be in the $J_e=3/2\rightarrow J_g=5/2$ transition of OVI whose Einstein coefficient is an order of magnitude below those of other transitions within the red wing of the blended feature and whose maximal polarization $p_0$ is 1.5\%. 

The high polarization of the SiIV$+$OIV blend leads us to conclude that resonant scattering is not the dominant polarization mechanism in SDSS~J652. In what follows we therefore only consider dust and electron scattering. 

\section{Model setup}
\label{sec:model}

\subsection{Overall geometry of the problem}
\label{sec:model_geometry}

In our model, both the emission-line region and the scattering region are axisymmetric. The model allows for a variety of conical morphologies for both, i.e., a polar scattering region, or an equatorial scattering region, or one that's confined to a narrow range of polar angles, and similar morphologies for the emission region. The key simplifications of our model are (i) that the emission-line region is point-like compared to the scatterer, (ii) that the velocity of the scatterer is purely radial and constant as a function of distance, (iii) that multiple scattering events are negligible, and (iv) that there is no extinction before or after scattering. Under these assumptions, each radial shell of the scatterer produces scattered light with the same kinematic pattern: the same kinematic structure of the scattered line, the same polarization fraction and the same polarization position angle. The impact of multiple scattering has been considered by \citet{lee97} and \citet{goos07}, and the effects of extinction within the scattering region by \citet{mari12}. Here we instead would like to specifically focus on the effects of outflow geometry and connect them to observations. 

If we are interested in the kinematic structure of the scattered emission and in the fractional polarization, but not in the overall scattered intensity, in our models with conical symmetry of scatterer we can consider only one radial shell. An additional integration of all Stokes parameters over the radial shells, each with its own geometry, would allow for any axisymmetric scattering structures. While we do not consider rotating winds here \citep{youn00}, the code can be amended to include axisymmetric rotation as well, by incorporating the rotational component of velocity into Doppler shift equations (\ref{eq:dopp1})-(\ref{eq:dopp3}). 

\begin{figure}
\includegraphics[scale=0.4, clip=true, trim=8cm 7cm 6cm 6cm]{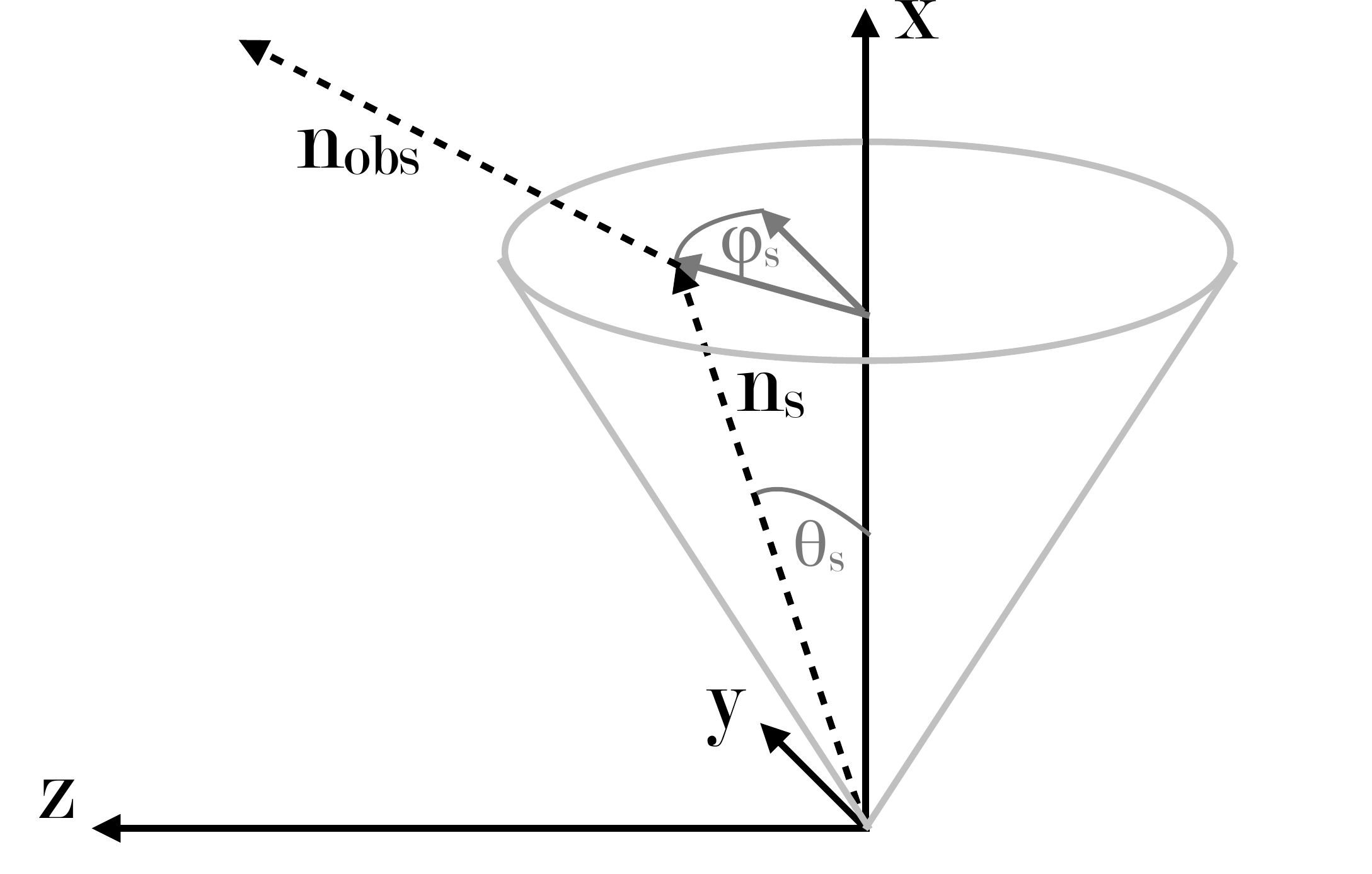}
\caption{The coordinate system associated with an axially symmetric scattering outflow. $x$ is the polar axis and the $x-z$ plane is parallel to the direction toward the observer. Polar angles $\theta$ are measured off the polar axis $x$ and azimuthal angles $\varphi$ are measured off the $y$-axis. A photon propagates from the compact central region at (0,0,0) in the direction of ${\bf n_{\rm s}}$ and is then scattered toward the observer. As seen by the observer, the $y$ axis is in the plane of the sky, perpendicular to the projected axis of symmetry.}
\label{pic:coord}
\end{figure}

The coordinate system is set up in Figure \ref{pic:coord}. The object's axis of symmetry is along the $x$ axis, and the observer is in the $x-z$ plane. $y$ axis is then added to produce a right-handed coordinate system $x-y-z$. To translate between spherical and Cartesian systems as necessary, the spherical system is set up with $x$ as its polar axis, polar angles $\theta$ are counted from the $x$-axis and azimuthal angles $\varphi$ are counted from the $y$-axis. 

In this coordinate system, the unit vector toward the observer is
\begin{equation}
{\bf n_{\rm obs}}=(\cos \theta_{\rm obs}, 0, \sin \theta_{\rm obs}).
\end{equation}
Photons originate in a compact emission region at the center of the coordinate system and propagate along the directions with unit vectors
\begin{equation}
{\bf n_{\rm s}}=(\cos\theta_{\rm s},\sin\theta_{\rm s}\cos\varphi_{\rm s},\sin\theta_{\rm s}\sin\varphi_{\rm s}).
\end{equation}
The scattering angle $\psi$ is the angle between the initial direction of propagation and the scattered direction, i.e., the direction toward the observer, so that
\begin{equation}
\cos\psi={\bf n_{\rm obs}}\cdot{\bf n_{\rm s}}. 
\end{equation}

\subsection{Phase function and polarization fraction}

Regardless of the scattering mechanism (electron scattering, dust scattering, resonant scattering), both the phase function of scattering -- i.e., the angular dependence of the scattering cross-section -- and the polarization fraction of scattered light depend only on the scattering angle $\psi$. We define the phase function through the differential cross-section as 
\begin{equation}
\frac{{\rm d}\sigma}{{\rm d}\Omega}=\frac{1}{4\pi} \sigma_{\rm total}g(\psi),
\end{equation}
so that $\int g(\psi){\rm d}\Omega=4\pi$. Both for electron (Thomson) scattering and for Rayleigh scattering (when the dust particles have sizes much smaller than the wavelength of light), the phase function of scattering is $g_{\rm TR}(\psi)=3(1+\cos^2\psi)/4$. 

Polarization fraction $p(\psi)$ defined here specifically for unpolarized incident light is a signed value $(I_{\perp}-I_{\parallel})/(I_{\perp}+I_{\parallel})$, where $I_{\perp}$ and $I_{\parallel}$ are the intensities of scattered light in polarization modes perpendicular and parallel to the scattering plane \citep{drai03}. Therefore, the sign of $p(\psi)$ tells us the dominant direction of the emerging polarization. In Thomson and Rayleigh scattering $p_{\rm TR}(\psi)=\sin^2\psi/(1+\cos^2\psi)\ge 0$, so the emerging polarization is always perpendicular to the scattering plane. 

Astrophysical dust contains particles of many different sizes, not necessarily small compared to the wavelength, so the Rayleigh approximation is insufficient. \citet{drai03} uses Mie approximation (spherical dust particles) to calculate the phase function $g_{\rm dust}(\psi)$ and the polarization fraction $p_{\rm dust}(\psi)$ for dust size distributions appropriate for the Milky Way and the Magellanic Clouds dust. We interpolate over the values in these numerical tables to obtain continuous phase and polarization functions. While $p_{\rm dust}(\psi)$ is almost always positive for dust scattering, it can be negative for obtuse scattering angles (backward scattering) for some combinations of size distributions and wavelengths, resulting in polarization within the scattering plane (the effect is best illustrated in Figure 5 of \citealt{drai03}). We retain the sign information in $p_{\rm dust}(\psi)$ to correctly incorporate it into the calculation of the polarization position angle. 

\subsection{Calculating the line-of-sight velocity distribution}

The point-like emission-line source has its own velocity structure, which would be blocked from the observer in an obscured (type 2) active nucleus but could be seen to an unobscured (type 1) observer. Following the model for Mrk~231 by \citet{veil16}, we consider line emission (e.g., CIV) to be isotropically produced by the gas outflowing with velocity $v_{\rm in}$ in an axisymmetric geometry, which is then scattered on much larger scales by a wind moving with a potentially different velocity $v_{\rm s}$ in a potentially different axisymmetric geometry. 

To calculate the Doppler shift resulting from scattering, we start with an emitter moving with ${\bf v_{\rm in}}$ which makes a photon propagating along ${\bf n_{\rm s}}$, which in turn hits the scatterer moving with ${\bf v_{\rm s}}$. Because we only consider radial motions for the scatterer, ${\bf n_{\rm s}}$ and ${\bf v_{\rm s}}$ are co-directional. In the rest-frame of the emitter, the photon has wavelength $\lambda_0$, the laboratory wavelength of the emission line in question. To the first order in $v/c$ (sufficient for winds with velocities $\sim$ a few thousand km s$^{-1}$) the scatterer sees this photon coming at it with 
\begin{equation}
\lambda'=\lambda_0\left(1+\frac{v_{\rm s}-{\bf v_{\rm in}}\cdot{\bf n_{\rm s}}}{c}\right).
\label{eq:dopp1}
\end{equation}
As seen in the scatterer frame, the photon arrives and is then scattered with the same wavelength. But the observer sees another Doppler shift due to the motion of the scatterer:
\begin{equation}
\lambda_{\rm obs}=\lambda'\left(1-\frac{{\bf v_{\rm s}}\cdot{\bf n_{\rm obs}}}{c}\right).
\label{eq:dopp2}
\end{equation}
Therefore, the observer would infer the line-of-sight velocity 
\begin{equation}
v_{\rm LOS}=v_{\rm s}(1-\cos\psi)-{\bf v_{\rm in}}\cdot{\bf n_{\rm s}}
\label{eq:dopp3}
\end{equation}
(positive for redshift and negative for blueshift). We assume that the cosmological redshifts have already been taken into account and that we are considering an observer which is at a large distance from the nucleus, but is in the rest frame of its host galaxy. 

To calculate the complete line profile of scattered intensity we integrate over the distribution function for each emitting and each scattering direction:
\begin{eqnarray}
I(v_{\rm LOS})=\int \sin\theta {\rm d}\theta {\rm d}\varphi \sin\theta_s {\rm d}\theta_s {\rm d}\varphi_s g(\psi) \times \nonumber \\
\delta\left(v_{\rm LOS}-v_{\rm s}(1-\cos\psi)+{\bf v_{\rm in}}\cdot{\bf n_{\rm s}}\right). \label{eq_int}
\end{eqnarray}
This is a four-dimensional integral over the polar and azimuthal directions of the initial gas velocity (producing the intrinsic nuclear line profile) and over the polar and azimuthal directions of the scatterer. The Dirac delta function indicates that for a given observed velocity $v_{\rm LOS}$ and given outflow velocities $v_{\rm in}$ and $v_{\rm s}$, there is only a small subset of angles in the parameter space $(\theta, \varphi, \theta_s, \varphi_s)$ that would result in this observed velocity. 

To calculate polarization, we use Stokes parameters, which have the advantage of being additive, unlike polarized intensity, which can cancel if we add up scattered beams with different polarization angles. To define Stokes parameters, we set the observer's projection of the $x$-axis on the plane of the sky to be astronomical North. The positive $y$-direction then becomes astronomical East and we can measure polarization position angles $\beta$ in their standard way to be East of North, with $\beta=0\deg$ if the electric field of the polarized light as seen by the observer is along the cone axis projected on the plane of the sky. If polarized intensity is $P$, then Stokes parameters are defined as
\begin{equation}
Q=P\cos2\beta; \,\,\,\, U=P\sin 2\beta. \label{eq_stokes}
\end{equation}
These parameters (defined relative to the projected axis which is taken to be the astronomical North) are directly comparable to the $Q'$ and $U'$ defined above for observational data. 

\begin{figure}
\includegraphics[scale=0.4, clip=true, trim=8cm 7cm 6cm 6cm]{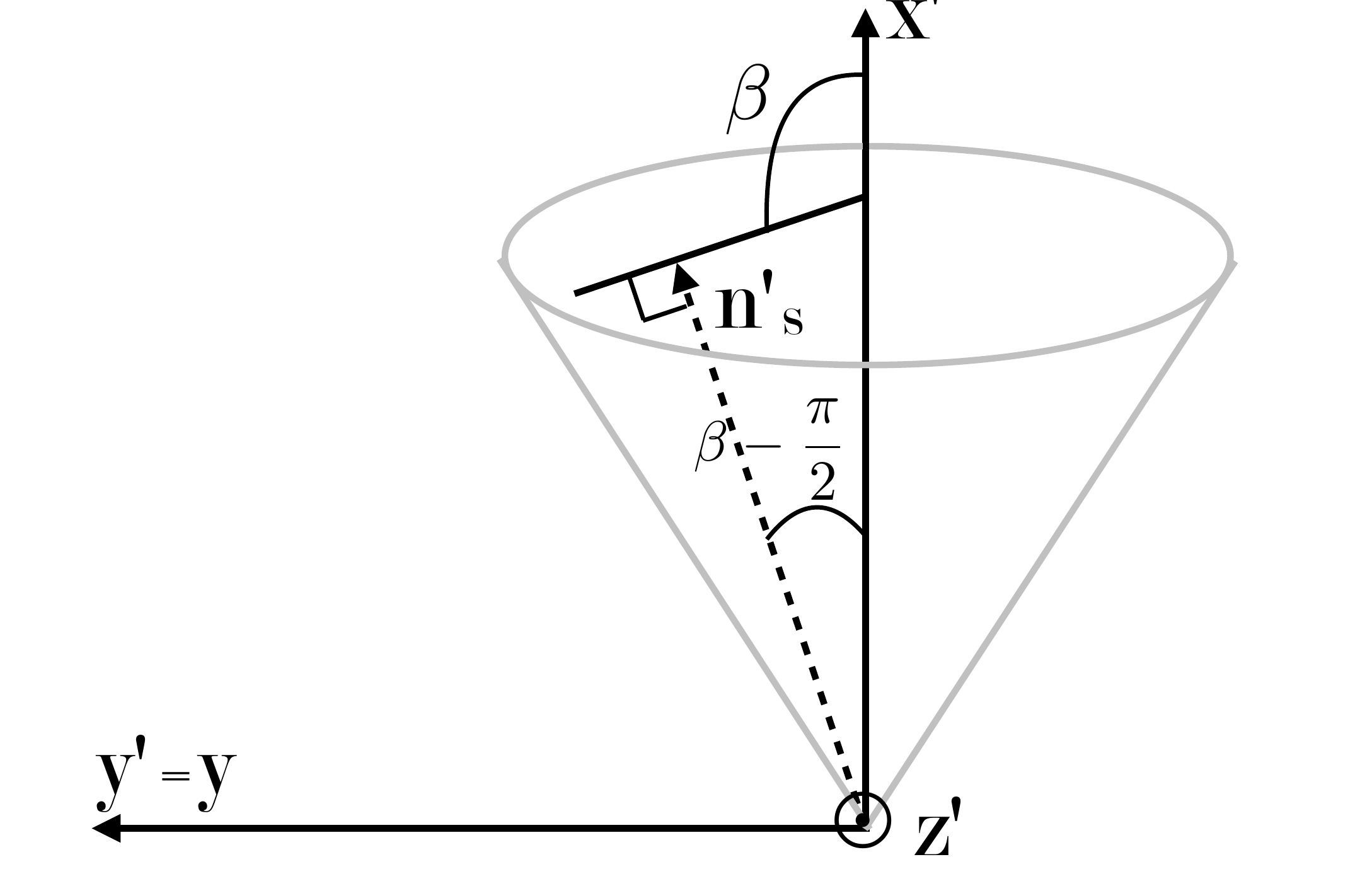}
\caption{The scattering event as seen by the observer in the plane of the sky. $x'$ is the projection of the axis of symmetry onto the plane of the sky, $z'$ is the direction toward the observer, and $y'=y$ axis is the same as the one in Figure \ref{pic:coord}. Once the light propagating with direction ${\bf n_{\rm s}}$ scatters off, in most cases it acquires polarization with position angle $\beta$ (polarization is perpendicular to the scattering plane), which can be calculated from the projection of ${\bf n_{\rm s}}$ onto the plane of the sky. The solid black line shows the dominant orientation of the electric field of the polarized light.}
\label{pic:geom}
\end{figure}

For most scattering processes, the polarization position angle of the electric field of the polarized light is perpendicular to the scattering plane (an interesting exception to this rule is discussed in Section \ref{ssec:polar}). But as the propagation vector ${\bf n_{\rm s}}$ sweeps its allowed directions, the orientation of the scattering plane changes (Figure \ref{pic:geom}). Therefore, to determine $\beta$ for every incidence of scattering, we need to measure the projection of ${\bf n_{\rm s}}$ onto the plane of the sky $x'-y'$. Applying a rotation transformation in the $x-z$ plane, we find the projected components to be 
\begin{eqnarray}
n'_x=\sin\theta_{\rm obs}\cos\theta_{\rm s}-\cos\theta_{\rm obs}\sin\theta_{\rm s}\sin\varphi_{\rm s};\nonumber \\
n'_y=\sin\theta_{\rm s}\cos\theta_{\rm s}.
\end{eqnarray}
Since Stokes parameters are additive, they can be summed up as they arise in different parts of the scattering region, so that the Stokes parameters of the line profiles are
\begin{eqnarray}
Q(v_{\rm LOS})=\int \sin\theta {\rm d}\theta {\rm d}\varphi \sin\theta_s {\rm d}\theta_s {\rm d}\varphi_s g(\psi) p(\psi)\times \nonumber \\
\left(-\frac{(n'_x)^2-(n'_y)^2}{(n'_x)^2+(n'_y)^2}\right)\delta\left(v_{\rm LOS}-v_{\rm s}(1-\cos\psi)+{\bf v_{\rm in}}\cdot{\bf n_{\rm s}}\right); \nonumber \\
U(v_{\rm LOS})=\int \sin\theta {\rm d}\theta {\rm d}\varphi \sin\theta_s {\rm d}\theta_s {\rm d}\varphi_s g(\psi) p(\psi)\times \nonumber \\
\left(-\frac{2n'_xn'_y}{(n'_x)^2+(n'_y)^2}\right)\delta\left(v_{\rm LOS}-v_{\rm s}(1-\cos\psi)+{\bf v_{\rm in}}\cdot{\bf n_{\rm s}}\right). \nonumber \\ \label{eq_pol}
\end{eqnarray}

Here $p(\psi)$ is again the polarization fraction, $g(\psi)$ is the same phase function of scattering (angular dependence of scattering) as the one used in the intensity profile, and the terms dependent on $n'_x$ and $n'_y$ are the same as the $\beta$ terms in equations (\ref{eq_stokes}). They account for the geometric dilution of polarization: if scattering occurs over a wide range of position angles of the incident photons as seen in the plane of the sky, even with high per-scattering polarization, the net polarization is lowered, until in the extreme case of centro-symmetric scattering the net polarization is zero. For dust scattering, expressions (\ref{eq_pol}) correctly take into account the unusual case of polarization sign reversal when $p_{\rm dust}(\psi)$ is used a signed function. 

With our notation, $Q$ is positive when polarization position angle is along the polar axis of the object. Therefore, for narrow cones where the scattering planes are close to $x-z$, net $Q$ values should be negative because the dominant direction of scattering is perpendicular to the scattering plane and parallel to the $y$ axis, with $\beta=90\deg$. Non-zero $U$ values reflect polarization orientation at 45 and 135 degrees to the cone axis. Because of the symmetry of our problem we expect $U$ to be zero, so we show it here only as a check on the accuracy of our calculations. 

The model is numerically implemented in \verb|Python|\footnote{\url{https://github.com/zakamska/polarized_outflows}}. The multi-dimensional integrals (\ref{eq_int}) and (\ref{eq_pol}) are calculated using a Monte Carlo integrator adapted from \verb|mcint.integrate|\footnote{\url{https://pypi.org/project/mcint/}}, with the Dirac delta function approximated by a Gaussian. To quickly explore the parameter space of our models, we use $10^4$ Monte Carlo trials with the Gaussian dispersion set to $0.01-0.05$ of the outflow velocity $v_{\rm s}$. Smaller widths result in noisier curves (and would therefore require more Monte Carlo steps to calculate the integral with a higher accuracy), but larger widths degrade the achievable velocity resolution. If higher quality profiles are desired, the width of the Gaussian may be decreased while the number of Monte Carlo trials is simultaneously increased. For the final plots presented in the paper we use $10^5$ Monte Carlo trials and Gaussian dispersion of $0.01v_{\rm s}$. 

For $\theta_{\rm obs}=90\deg$, i.e., the system seen edge-on, it is enough to compute the Stokes parameters from only one side ($\theta_{\rm s}\le 90 \deg$) because the other side contributes equal $I$, $Q$ and $U$ and therefore does not affect the calculated polarization kinematics. If the observer is at $\theta_{\rm obs}<90\deg$, then in principle the two sides of the outflow should be considered separately and then their Stokes parameters added together, but in practice we only take into account the approaching side on the assumption that the back side of the outflow is heavily obscured. 

Continuum polarization can be obtained from equations (\ref{eq_int}) and (\ref{eq_pol}) by integrating over $v_{\rm LOS}$ (eliminating the delta function). The incident emission is comprised both of lines and continuum, and their ratio is a free parameter of the model. We take advantage of the additive nature of the Stokes parameters to calculate the overall scattered intensity, polarization fraction and position angle for the line + continuum combination.

\section{Model results and comparison with observations}
\label{sec:analysis}

In this Section, we explore a few example geometries using our model, explain some of the phenomena that arise in the resulting model profiles, and compare the model profiles to observations. In Section \ref{ssec:polar} we investigate the case of polar outflows with electron scattering, and in Section \ref{ssec:eq} equatorial and thick disk skin outflows with electron scattering. In Section \ref{ssec:elec} we discuss similarities and differences between electron and dust scattering. 

\subsection{Polar scattering outflow}
\label{ssec:polar}

In most our setups, the nuclear line profile is created by a point-like source, in which the emitting gas moves radially with velocity $v_{\rm in}$ and uniformly fills a cone with $\theta=0-\theta_{\rm max}$. None of the main qualitative results depends sensitively on this choice and in principle another model for the source can be implemented (e.g., one that includes a more realistic distribution of velocities) at the expense of more computational complexity since one would then have to integrate over the distribution of $v_{\rm in}$ values as well.  

In our first model -- polar scattering outflow observed edge-on -- this emission is then scattered into the line of sight by the larger-scale wind moving within a range of $\theta_{\rm s}=0-\theta_{\rm max}$ and with the same velocity $v_{\rm s}=v_{\rm in}$. We then envision a type 2 AGN observed edge-on, with $\theta_{\rm obs}=90\deg$. The nuclear line profile is obscured, so the spectrum that we see is entirely due to the scattered light, part of which is polarized. 

\begin{figure}
\includegraphics[scale=0.7]{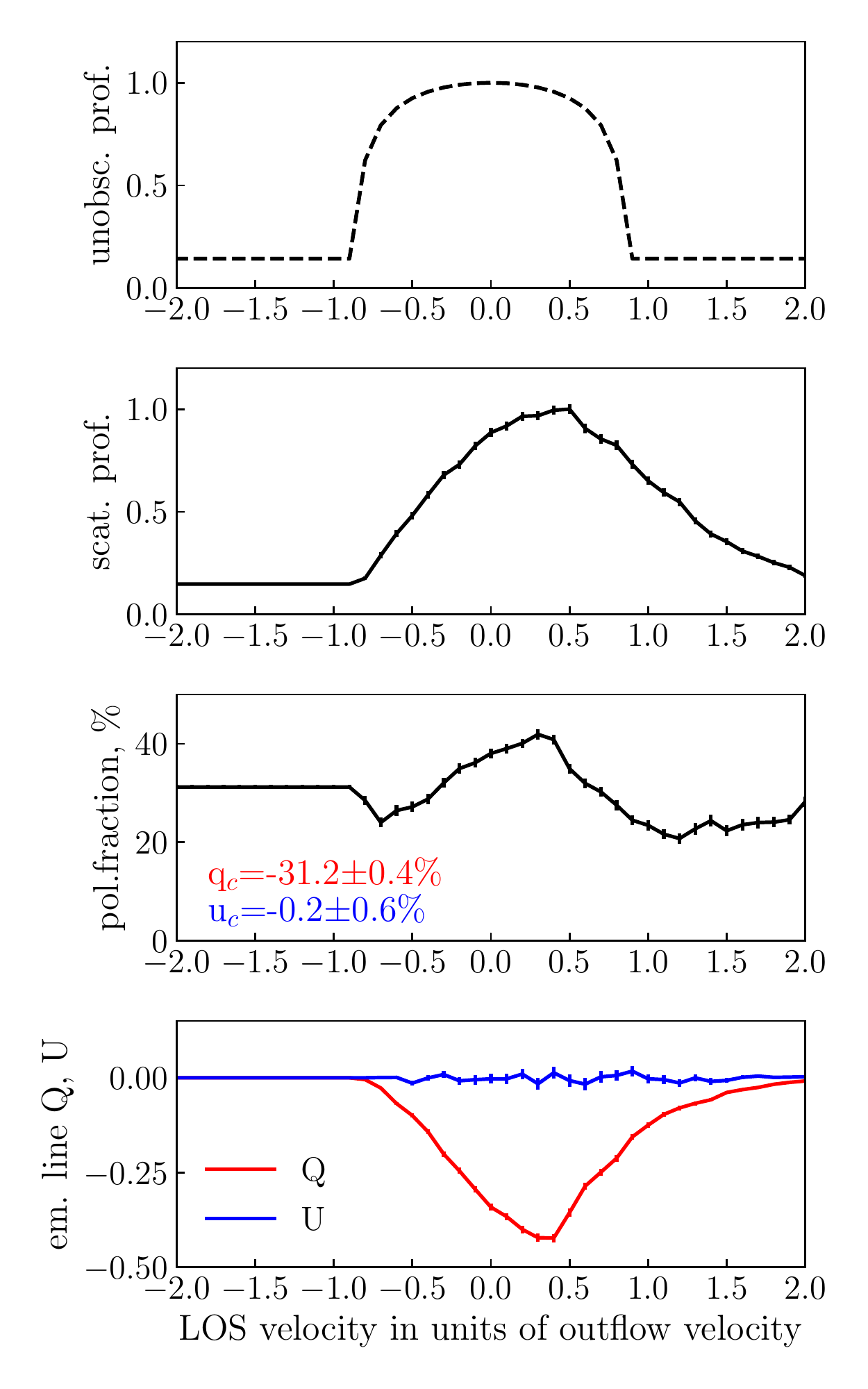}
\caption{Results for a polar emitter ($\theta=0-\theta_{\rm max}$) and a polar scatterer ($\theta_{\rm s}=0-\theta_{\rm max}$) with $v_{\rm in}=v_{\rm s}$ and $\theta_{\rm max}=60\deg$ viewed edge-on ($\theta_{\rm obs}=90\deg$). From top to bottom: the nuclear spectrum (emission line plus continuum), the scattered light spectrum (emission line plus continuum), polarization fraction of the scattered light spectrum (emission line plus continuum), with continuum fractional polarization $q_{\rm c}$ and $u_{\rm c}$ indicated on the label, and $Q$ and $U$ Stokes intensities for the emission line alone. The error bars reflect the Monte Carlo sampling of the multi-dimensional integrals (\ref{eq_int}) and (\ref{eq_stokes}). The projected polar axis of the object defines the Northern direction. $Q$ is negative across the emission line and the continuum, so that the polarization position angle of this spectrum is perpendicular to the outflow axis. The total line profile and the scattered line profile are normalized to unity at peak; polarization fraction is as observed by the observer; and $Q$ and $U$ intensities are in the same units as the scattered profile.}
\label{pic:polar}
\end{figure}

In Figure \ref{pic:polar} we present the results of calculations for $\theta_{\rm max}=60\deg$ and a phase function and polarization fraction appropriate for electron scattering. The line-of-sight velocity profile that would have been seen in the absence of obscuration is shown in the top panel. This profile can be calculated by analogy to equation (\ref{eq_int}): $I_{\rm unobsc}(v_{\rm LOS})=\int \sin\theta {\rm d}\theta{\rm d}\varphi \delta(v_{\rm LOS}+{\bf v_{\rm in}}\cdot{\bf n_{\rm obs}})$ and for $\theta_{\rm obs}=90\deg$ can be calculated analytically: $I_{\rm unobsc}(v_{\rm LOS})=2\arccos\left(v_{\rm in}\cos\theta_{\rm max}/\sqrt{v_{\rm in}^2-v_{\rm LOS}^2}\right)/v_{\rm in}$ for $|v_{\rm LOS}|<v_{\rm in}\sin\theta_{\rm max}$ and 0 otherwise. Here $\theta_{\rm max}\le 90\deg$; the cases of $\theta_{\rm max}>90\deg$ or equatorial emitters can be reduced to the linear combinations of $I_{\rm unobsc}$. 

Since Stokes parameters are additive, we can add $I$, $Q$ and $U$ values for the scattered emission line and the scattered continuum to obtain the total observed spectrum (2nd panel) and its polarization fraction (3rd panel). Compared to the profile that would have been seen directly from the emitter, the scattered line profile (2nd panel) shows a redshifted tail. This part of the spectrum originates from the back part of the outflow which is redshifted away from both the observer and the emitter, combining the Doppler effects. 

\begin{figure*}
\includegraphics[width=\textwidth]{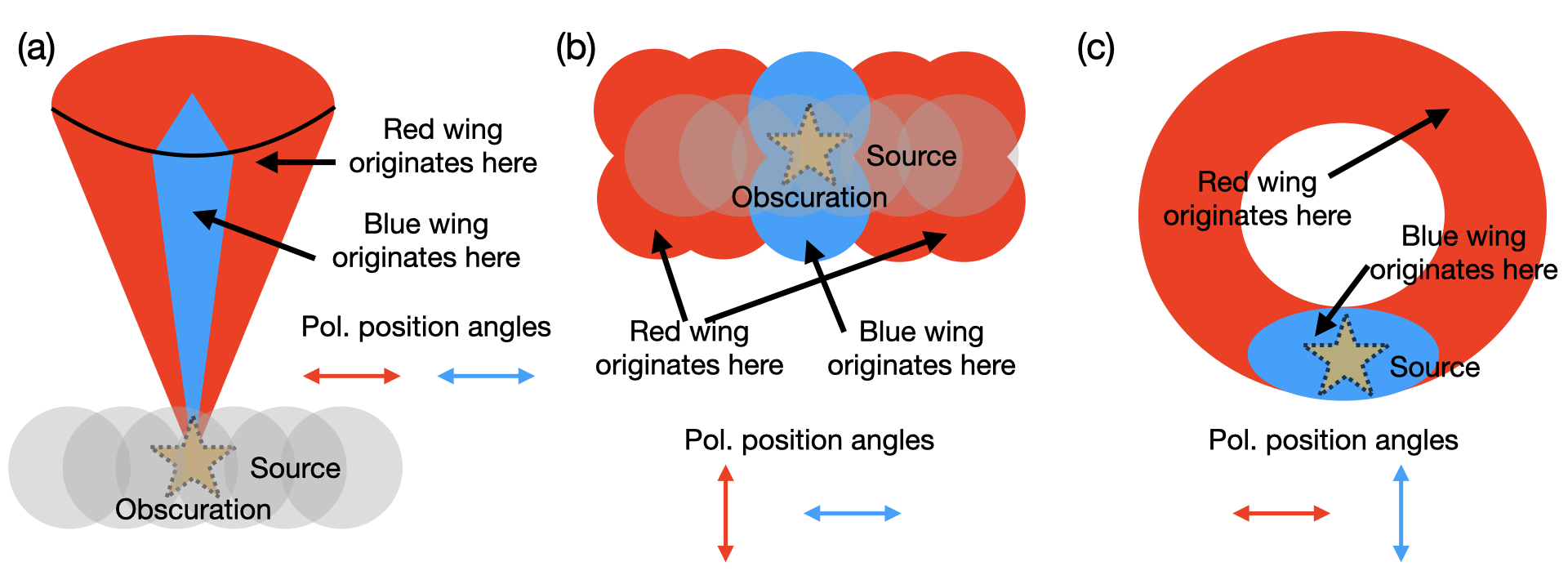}
\caption{Overall orientation of the scatterer in the plane of the sky and the resulting polarization position angles as seen by the observer. The polar axis of the object is vertical in all cases. (a) For a filled scattering cone, the net projected orientation of the scatterer across all parts of an emission line is along the axis, so the resulting polarization position angles are perpendicular to the projected axis. (b) In an equatorial outflow observed edge-on, most of the scatterer is perpendicular to the axis, resulting in the polarization position angle along the axis. However, for thick enough equatorial outflow the scatterer responsible for the blue wing of the emission line can be elongated along the axis, resulting in a polarization position angle swing. (c) The scattering outflow proceeds within a range of polar angles, which includes the observer's line of sight. This geometry can result in a polarization position angle swing and reproduce other features of SDSS~J1652.}
\label{pic:obs}
\end{figure*}

In the bottom panel we show the Stokes intensities for the emission line alone, without the continuum. Due to the symmetry of the problem, the $U$ Stokes intensity is supposed to be exactly zero in our model, and it is being displayed only as a check on the calculations and to demonstrate the accuracy of the numerical integration in equations (\ref{eq_pol}). The $Q$ Stokes intensity is negative across the entire emission-line profile, and so is its velocity integral which is proportional to the continuum polarization. Therefore $Q$ for the total line+continuum spectrum is negative everywhere, so that the polarization position angle -- as expected -- is orthogonal to the symmetry axis as seen in the plane of the sky (Figure \ref{pic:obs}a). 

\begin{figure}
\includegraphics[scale=0.7]{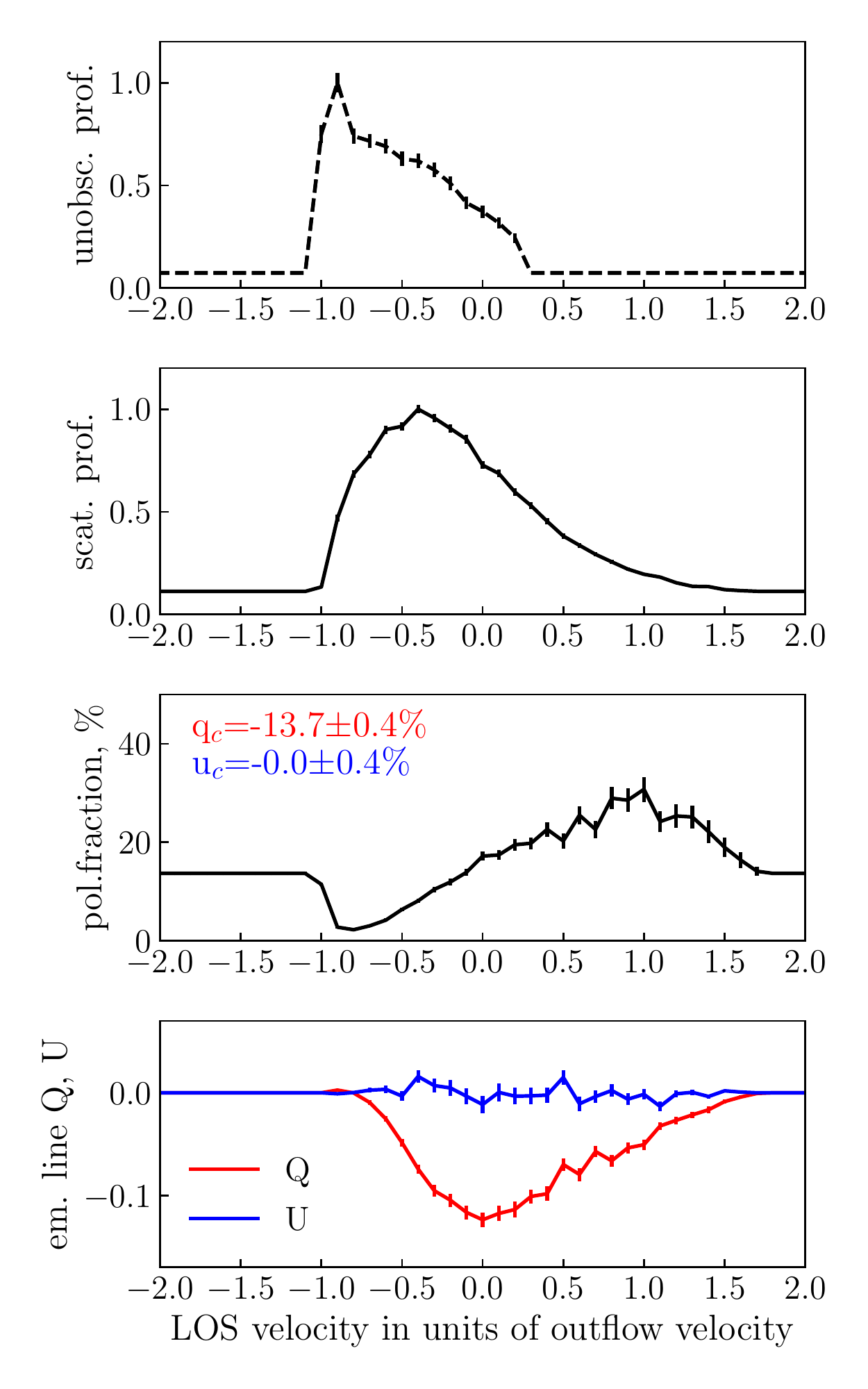}
\caption{Results for a polar emitter and a polar scatterer with $\theta_{\rm max}=60\deg$ viewed within the outflow at $\theta_{\rm obs}=45\deg$. From top to bottom: the nuclear spectrum (emission line plus continuum), the scattered light spectrum (emission line plus continuum), polarization fraction of the scattered light spectrum (emission line plus continuum), with continuum fractional polarization $q_{\rm c}$ and $u_{\rm c}$ indicated on the label, and $Q$ and $U$ Stokes intensities for the emission line alone. The net values of polarization are significantly lower than in the edge-on case, the polarized fraction shows interesting kinematic structure on the blue side, and the peak of the polarized line is offset to the red by comparison to the peak of the scattered emission.}
\label{pic:polar_within}
\end{figure}

In Figure \ref{pic:polar_within} we show the same polar scatterer, but now viewed along a line of sight within the outflow. The net level of polarization is now significantly lower than in the edge-on case: this is due both to the smaller polarization fraction $p_{\rm TR}(\psi)$ for forward-scattering and to the partial geometric cancellation of polarization (in the extreme case of on-axis view the polarization is exactly zero). The profiles now display some of the interesting features we highlighted for SDSS~J1652: the polarization fraction dips on the blue side of the line and the polarized profile is redshifted by comparison to the scattered profile by about 0.5$v_{\rm s}$. Both of these effects are due to the fact that the blue wing of the polarized line is formed by the most forward-scattering part of the outflow where $p_{\rm TR}$ is nearly zero. $Q$ values for the continuum and for the line are negative (meaning the polarization position angle is orthogonal to the projected axis of symmetry); this is also consistent with observations of SDSS~J1652. However, the tantalizing `swing' of the polarization position angle does not appear in the polar model regardless of the observer's orientation. 

\subsection{Equatorial scattering outflow}
\label{ssec:eq}

Equatorial dusty winds lifted off the obscuring material surrounding the AGN have been proposed on the basis of many observations \citep{will92, veil16}, as well as by theoretical work \citep{koni94, elit06, chan16}. In this section we consider the emitter to be expanding within a filled cone (polar emitter, as before), which determines the unobscured velocity profile, but the scatterer is now in an outflow confined between angles $\theta_{\rm min}$ and $\theta_{\rm max}$. If $\theta_{\rm max}$ for this outflow is close to $90\deg$, then such outflow would be reasonably called `equatorial'. Another situation of astrophysical interest is that of a `disk skin' outflow confined between two angles significantly smaller than $90\deg$, it is relevant for outflows along the surface of geometrically thick disks.

\begin{figure}
\includegraphics[scale=0.7]{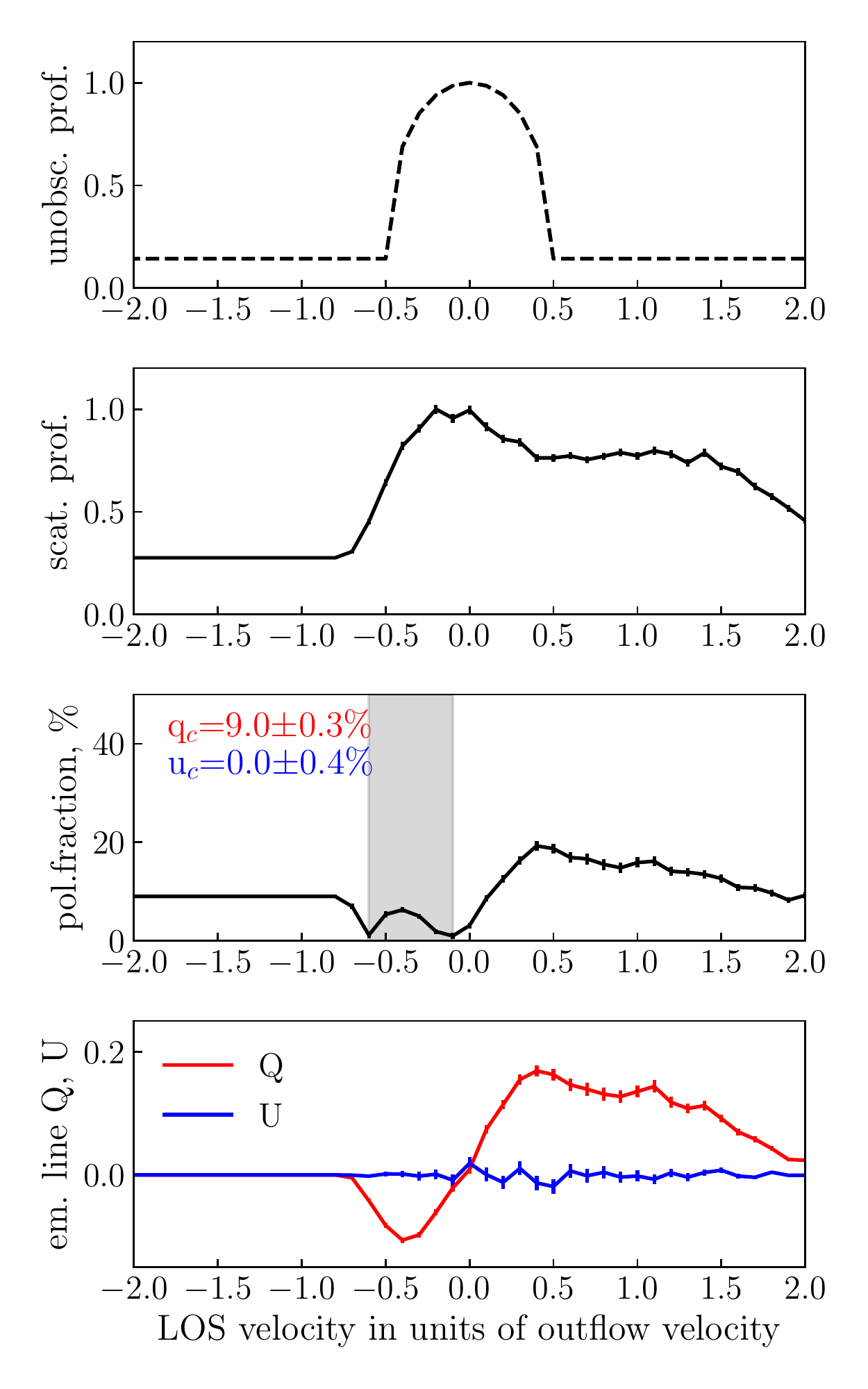}
\caption{Model spectropolarimetry for a polar emitter ($0<\theta<30\deg$) and an equatorial scatterer ($30<\theta_{\rm s}<90\deg$) with $v_{\rm in}=v_{\rm s}$. From top to bottom: the nuclear spectrum (emission line plus continuum), the scattered light spectrum (emission line plus continuum), polarization fraction of the scattered light spectrum (emission line plus continuum), and $Q$ and $U$ Stokes intensities for the emission line alone. $Q$ changes sign across the emission line, even with the addition of the continuum. The net result is that the polarization position angle is aligned with the axis everywhere $(Q>0)$ except the shaded part of the velocity profile, where the polarization position angle swings by 90$\deg$. }
\label{pic:eq}
\end{figure}

The model with an emitter expanding with $0<\theta<30\deg$ and a scatterer expanding with $30<\theta_{\rm s}<90\deg$ (Figure \ref{pic:eq}) is at first glance quite promising for explaining several observed features of SDSS~J1652: (i) the redshifting of the peak of the polarized line profile relative to the peak of the scattered profile; (ii) a $90\deg$ swing of the polarization position angle; (iii) the net level of polarization within the line is lower on the blue side of the line and higher on the red side of the line.

In this realization of the model, the swing of the polarization angle is due to the projected orientation of the scatterer as seen by the observer in different parts of the emission line. The blue side of the scattered line comes from the part of the outflow moving toward the observer. Given a sufficient thickness of the equatorial outflow, the gas producing the blueshifted part of the outflow may end up with an orientation along the axis of symmetry, so we expect negative $Q$ for this part (Figure \ref{pic:obs}b). In contrast, the red side of the scattered line comes from the sides which have a large scattering volume and are moving away from the emitter resulting in redshifted scattered emission. We expect positive $Q$ values for this part of the outflow, as well as for the continuum. The parameter space for the `swing' to occur in the edge-on view is somewhat limited: a thinner `skin' outflow with $30<\theta_{\rm s}<50\deg$ behaves like a polar scatterer in Figure \ref{pic:polar} with $Q<0$ everywhere across the emission line, and a thinner equatorial outflow with $60<\theta_{\rm s}<90\deg$ has $Q>0$. 

The key observable in SDSS~J1652 which contradicts the model in Figure \ref{pic:eq} is the orientation of the net polarization: the position angle of the projected axis of SDSS~J1652 is known from direct imaging and integral-field observations, and it is well-measured that the polarization of the continuum and of the red wing of the line are orthogonal to that axis (negative $Q'$ values in Figure \ref{pic:1652_lines}), whereas the model in Figure \ref{pic:eq} unsurprisingly predicts that the net polarization should be aligned with the projected axis (positive $q_{\rm c}$ values and $Q>0$ on the red wing).

\begin{figure}
\includegraphics[scale=0.7]{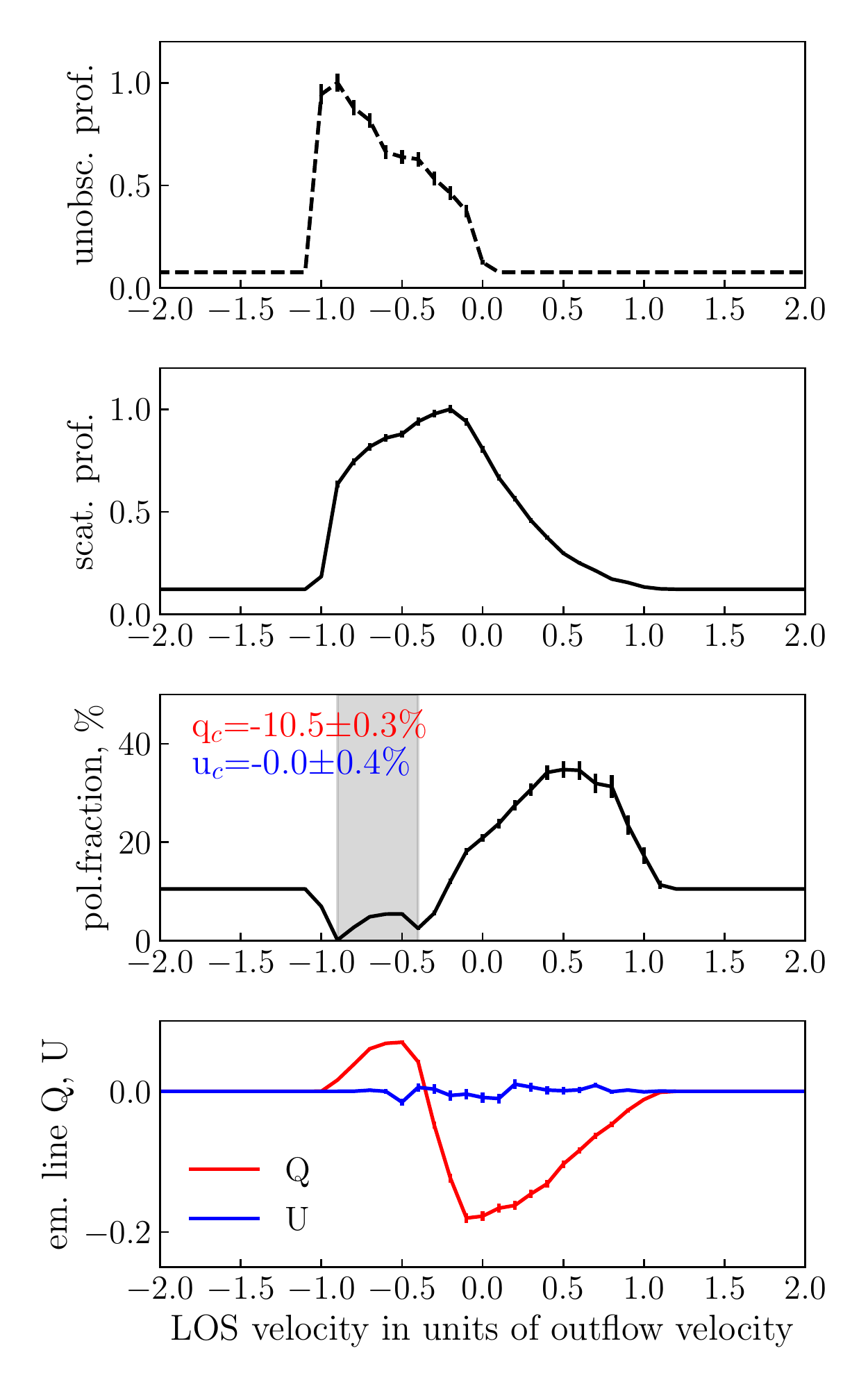}
\caption{Model spectropolarimetry for a polar emitter ($0<\theta<50\deg$) and a `skin' scatterer ($30<\theta_{\rm s}<50\deg$) viewed within the outflow at $\theta_{\rm obs}=40\deg$. From top to bottom: the nuclear spectrum (emission line plus continuum), the scattered light spectrum (emission line plus continuum), polarization fraction of the scattered light spectrum (emission line plus continuum), and $Q$ and $U$ Stokes intensities for the emission line alone. The `swing' of the sign of $Q$ from positive to negative, the net negative values of $Q$ for the continuum and the red wing of the line, the reduced polarization fraction on the blue wing and the redshift of the polarized profile relative to the scattered profile are all in agreement with the features seen in SDSS~J1652. }
\label{pic:eq_within}
\end{figure}

Our last class of geometries shown in Figure \ref{pic:eq_within} is the `skin' equatorial outflow viewed close to or through the outflow. This model qualitatively matches all of the features of line polarization we highlighted in SDSS~J1652 in Section \ref{sec:1652}. It reproduces the `swing' of the polarization position angle with the signs of $Q$ in agreement with those seen in SDSS~J1652 (going from positive on the blue  side to negative on the red side and in the continuum). The projected geometry responsible for these orientations as viewed in the observer's plane is illustrated in Figure \ref{pic:obs}c. Mixing $Q<0$ continuum values with $Q>0$ blue wing values results in a low polarization fraction on the blue wing of the line, as observed. The polarized flux is redshifted by about $v_{\rm s}$ relative to the scattered flux. 

All these qualitative features are retained within some range of assumed parameters (e.g., $50<\theta_{\rm s}<70\deg$ and $20<\theta_{\rm s}<30\deg$ models show all these features as well, as long as the observer is within the $\theta_{\rm s}$ range), with quantitative changes in the net level of polarization (lower for smaller $\theta_{\rm s}$) and velocity structure (smaller velocity range for smaller $\theta_{\rm s}$). Thus, the `skin' outflow models viewed within the outflow are our primary models of interest for SDSS~J1652. 

\subsection{Effects of the phase function and of the polarization function}
\label{ssec:elec}

The phase function $g_{\rm dust}(\psi)$ and the polarization fraction $p_{\rm dust}(\psi)$ for dust are sensitive to the dust size distribution \citep{drai03}. Qualitatively, dust is strongly forward-scattering, with cross-section for scattering being over an order of magnitude higher at $\psi=0\deg$ than at $180\deg$. At a fixed dust size distribution, scattering cross-section decreases as the wavelength increases. For the Magellanic Clouds and for the Milky Way dust size distributions, the polarization curve is qualitatively similar to $p_{\rm TR}(\psi)$ at wavelengths $\ga 6000$\AA\ and $\la 1400$\AA, though peaking at $20-90\%$ depending on the wavelength instead of at $100\%$ for $\psi=90\deg$ as is the case for Thomson scattering. At intermediate wavelengths, $p_{\rm dust}(\psi)$ is very sensitive to the size distribution of the particles, it becomes dissimilar from $p_{\rm TR}(\psi)$ and peaks at values $<20\%$.  

Another distinct feature of dust scattering in \citet{drai03} is that at large scattering angles $\psi\sim 150\deg$ the polarization position angle can be in the scattering plane, resulting in $p_{\rm dust}<0$, which is in contrast to Rayleigh, Thomson and resonant scattering which always results in polarization perpendicular to the scattering plane. This polarization reversal has been studied in the context of back-scattered light from comets and debris disks, but usually with dust agglomerates (\citealt{shen09} and references therein). Nonetheless, the reversal is present in Mie theory when purely spherical dust grains are considered, due to interference effects that arise when the particle size is comparable to the wavelength of light \citep{wein01, drai03}, so we explore whether this reversal can cause any qualitative changes in the observed spectropolarimetry of quasar outflows.

\begin{figure}
\includegraphics[scale=0.7]{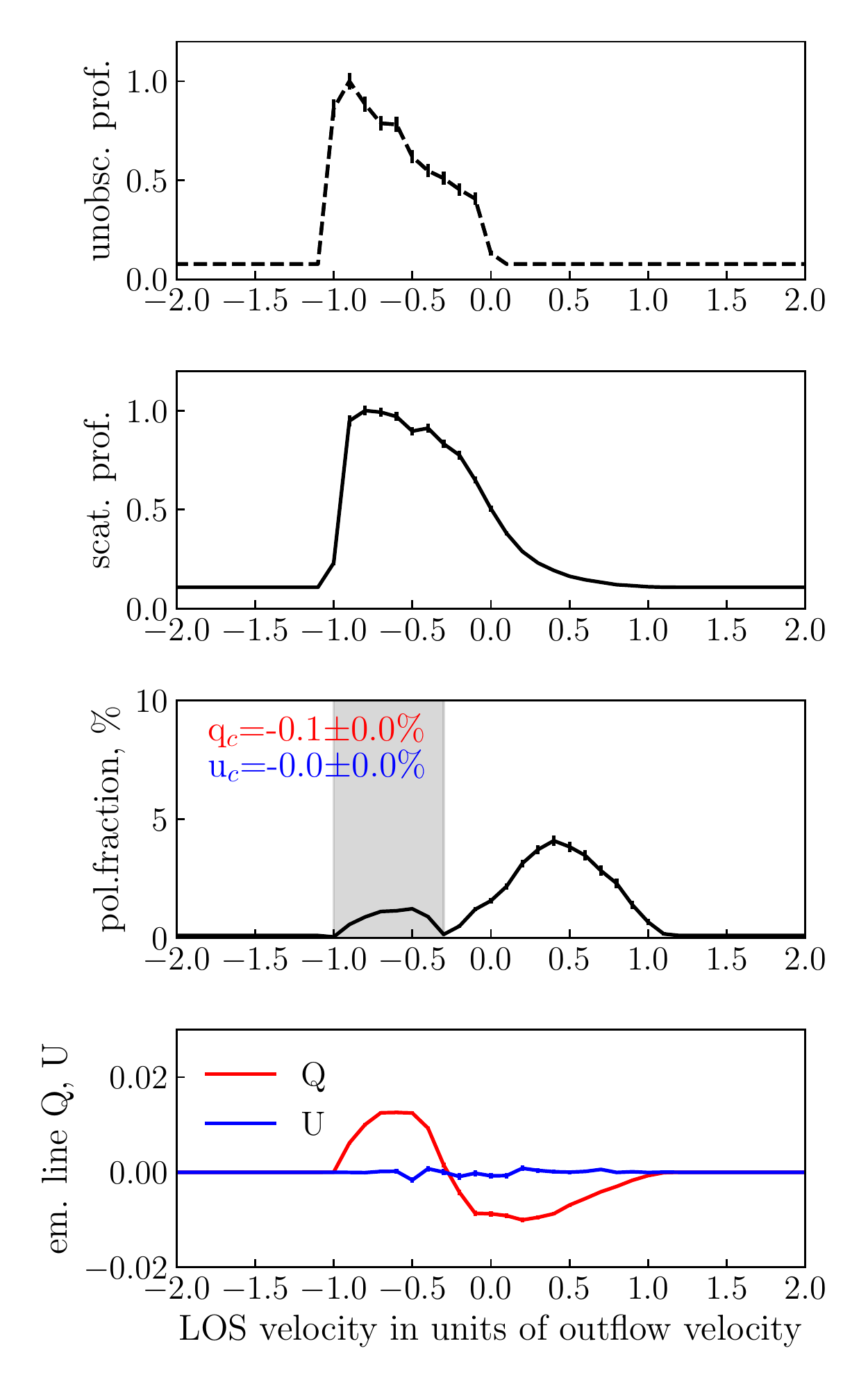}
\caption{Model spectropolarimetry for a polar emitter ($0<\theta<50\deg$) and a `skin' scatterer ($30<\theta_{\rm s}<50\deg$) viewed within the outflow at $\theta_{\rm obs}=40\deg$, now with dust phase function and polarization fraction curves. This is the same geometry of the outflow as that in Figure \ref{pic:eq_within}. While some of the key features (such as polarization position angle swing) are retained, the net polarization is significantly lower than in the electron-scattering case and well below the observed values.}
\label{pic:dust}
\end{figure}

We use the Small Magellanic Cloud (SMC) phase function $g_{\rm dust}(\psi)$ and polarization fraction $p_{\rm dust}(\psi)$ curves at 4685\AA\ from \citet{drai03}. While this is not the correct wavelength for our particular observation, we keep in mind that the dust size distribution in quasar outflows is unknown and can be quite distinct from those available in \citet{drai03}. For example, \citet{gask04} present extinction curves characteristic of quasar nuclear regions which are well explained by the paucity of small grains which are easily destroyed. We use the SMC curves at 4685\AA\ because they have a relatively high polarization fraction peak at 25\%, while simultaneously showing the polarization sign reversal described above. In Figure \ref{pic:dust}, we show the results for dust scattering by the outflow with the same geometry as our best `skin outflow' model shown in Figure \ref{pic:eq_within} with electron scattering. Our main feature of interest -- the swing of the polarization angle -- is still apparent, since it is largely due to the geometry of the projected scattering regions producing different parts of the emission lines (Figure \ref{pic:obs}c). 

The biggest issue with dust scattering models is that the net polarization is significantly lower than that in electron scattering models, to the point that they are in tension with observed values which reach 20\% in SDSS~J1652. This is especially true for the continuum: the scattered continuum is produced primarily by gas moving closest to the line of sight to the observer since the scattering efficiency is much higher for forward scattering, yet the polarization fraction is very small at these angles. We have tried a variety of plausible dust scattering curves and geometries, and in none of the combinations does the polarization fraction exceed 10\%, and in most it is below 5\%. Another related trend is that dust-scattered emission lines tend to be more blueshifted relative to the intrinsic profiles and narrower than electron-scattered ones because of the low efficiency of scattering from the parts of the outflow receding from the observer. 

The polarization fraction sign reversal for backward scattering can produce a swing in the polarization position angle. Indeed, for a polar emitter $0<\theta<80\deg$ and equatorial scatterer $80<\theta_{\rm s}<90\deg$ viewed edge-on ($\theta_{\rm obs}=90\deg$), the electron-scattering model predicts purely positive $Q$ values across the entire line profile. In contrast, the dust-scattering model shows negative $Q$ values at $v_{\rm LOS}\ga 2v_{\rm s}$, i.e., on the reddest part of the line profile which is produced by the part of the outflow receding from the observer and scattering at obtuse angles. Therefore, in principle this peculiarity of the dust polarization curves can result in the swing of the polarization position angle across the emission lines. But the effect is quite weak because backward scattering by dust is inefficient compared to the forward scattering, so the back of the polar outflow is barely visible -- and in practice would be even less so because of the intervening extinction. The polarization angle swing is not seen in the full line + continuum profile due to the addition of the positive Stokes parameters for the continuum. The polarization position angle of the continuum is matched on the blueshifted side of the velocity profile and swings on the redshifted part, whereas the opposite is observed in SDSS~J1652, so this particular model for the swing is not a good match for SDSS~J1652. Finally, the allowed range of geometries for this effect is very narrow: thicker equatorial outflows result in geometric polarization position angle swings as shown in Figure \ref{pic:eq} and Figure \ref{pic:obs}b. 

In conclusion, although the polarization position angle swing due to peculiarities of the dust polarization curve is a potentially interesting feature, it does not provide a plausible explanation for the observed polarization of the ``swinging'' objects due to the low scattering efficiency at the relevant obtuse scattering angles. Instead of relying on peculiarities of scattering and polarization curves to reproduce the swing, we need to rely on the geometry of the outflow. Furthermore, the low net polarization of the dust-scattered lines and continuum is in appreciable tension with the values observed in SDSS~J1652. This feature of dust scattering is difficult to eliminate by adjusting the particle size distribution due to the forward-scattering nature of dust: it is the forward-scattered light that dominates the scattered signal, but the fractional polarization is at its lowest for these angles. 

\section{Discussion}
\label{sec:disc}

\subsection{Geometric unification of high accretion rate sources}
\label{sec:dis_unif}

Polarization position angle swings \citep{smit05, cape21} -- appear quite common in type 1 quasars; the first recorded appears to be that in 3C382 by \citet{anto84}. In these sources, spectropolarimetry helps reveal the kinematic structures within the broad-line region: the scatterer appears to be equatorial and may be rotationally supported \citep{youn07}. Here we present a different kind of models for polarization swings -- models in which the scatterer is dominated by the outflow activity, without appreciable rotation. Thanks to the high levels of polarization and exquisite data, we can place stronger geometric constraint in the case of SDSS~J1652 specifically. In particular, the polarization position angle within the emission line and the orientation of the large-scale scattered-light nebula (which directly gives us the orientation of the projected symmetry axis) can only be reconciled if the observer's line of sight is within the outflow and if the outflowing material is distributed within a relatively narrow range of polar angles (Figure \ref{pic:eq_within}). 

\begin{figure}
\includegraphics[width=9cm]{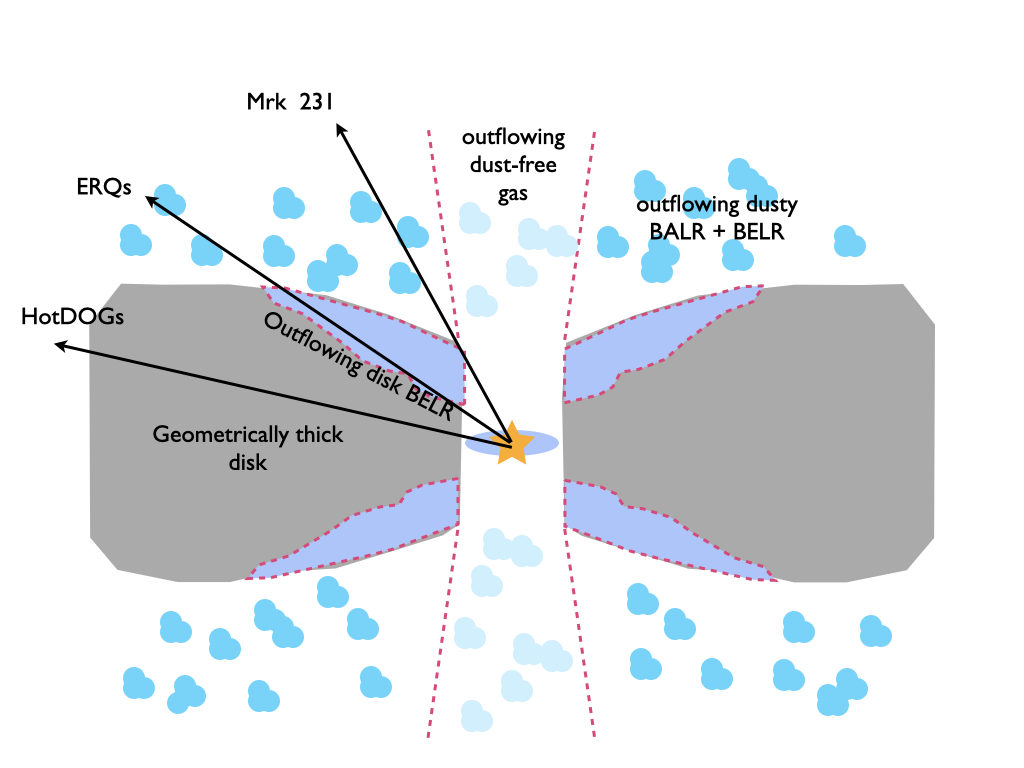}
\caption{The proposed unification of obscuration and polarization properties in several types of high accretion rate sources (adapted from \citealt{veil16} with permission). The model was developed to explain multi-wavelength properties and polarization of Mrk~231 \citep{veil13b, veil16}. Here we propose, based on observations of SDSS~J1652, that extremely red quasars (ERQs) are seen through the outflowing skin of their geometrically thick obscuring material, whereas hot dust-obscured galaxies (HotDOGs) are viewed closer to edge-on. BALR and BELR stand for broad absorption-line region and broad emission-line region, correspondingly.}
\label{pic:cartoon}
\end{figure}

This leads us to propose a unification model of high accretion rate sources -- such as Mrk~231 and SDSS~J1652 -- shown in Figure \ref{pic:cartoon}. Eddington ratios for these sources are not well known due to the difficulty of obtaining dynamical masses or applying standard scaling relationships to sources with strong outflows, but estimates range between 0.5$-$5 \citep{veil16, zaka19}, although at black hole masses that may differ by up to two orders of magnitude (from a few$\times 10^7M_{\odot}$ in the case of Mrk~231 to a few$\times 10^9M_{\odot}$ in ERQs). Multiple lines of evidence, including spectropolarimetry, suggest that both Mrk~231 and SDSS~J1652 are viewed through outflowing material, but rest-frame optical spectra and the overal spectral energy distributions of ERQs are significantly different from those of Mrk~231: ERQs have higher infrared-to-optical ratios \citep{zaka19} and rarely show evidence for circumnuclear emission of FeII \citep{zaka16b, perr19} characteristic of Mrk~231. This leads us to suggest that the line of sight in ERQs lies closer to the equatorial plane than it does in Mrk~231 (Figure \ref{pic:cartoon}), making them on average more dust-obscured. The column densities measured from X-ray observations in ERQs in general \citep{goul18a} and in SDSS~J1652 specifically \citep{ishi21} are in the Compton-thick regime ($N_H\sim 10^{24}$ cm$^{-2}$), which at face value would correspond to a visual extinction of $A_V=45$ mag \citep{ricc17} and prevent us from seeing any UV/optical emission, but this column density can be reconciled with our UV/optical data and Figure \ref{pic:cartoon} if the emission-line region is co-spatial with the outflow \citep{veil16} and is distributed on much larger spatial scales than the compact X-ray emitting region. 

Another intriguing and closely related population are hot dust-obscured galaxies (HotDOGs). These objects are selected to be near-infrared dropouts \citep{eise12}, and as a result they have even more extreme infrared-to-optical ratios than ERQs at comparable bolometric luminosities \citep{asse15} and at implied near- or super-Eddington accretion rates \citep{wu18}. Despite the extremely high levels of obscuration suggested by these spectral energy distributions \citep{ricc17}, some rest-frame UV emission is detected well in excess of any expected direct emission from the nucleus. It has now been confirmed by direct imaging and polarimetric observations that some of the light escapes from the nuclear regions along the polar opening in the obscuring material, scatters off and reaches the observer \citep{asse20, asse22}. The multi-wavelength properties of these sources are therefore well explained by a geometry similar to that suggested for ERQs, but with lines of sight that are closer to edge-on (Figure \ref{pic:cartoon}). %This might also be a good geometry to explain the spectropolarimetric properties of SDSS~J1623+3122, an obscured / type 2 quasar whose UV emission lines `swing' on the red side.

\citet{will92} and \citet{hine95} proposed that a model similar to that shown in Figure \ref{pic:cartoon} can explain some of the phenomenology of the broad absorption line (BAL) quasars, if the BAL features arise due to passage through the outflowing `skin' of the dusty torus. What is new in the data and models presented here is that the same region can now be identified with the scattering region in SDSS~J1652, and therefore the kinematic structure of the polarized emission lines can be used to probe the geometry and physical conditions of this region.  

It has long been known from analytical calculations that high accretion rates onto black holes are associated with geometrically thick disks (also known as `slim' disks, \citealt{abra88}). This is now firmly established by many numerical simulations which take into account the metric near the black hole and the dynamical effects of radiation \citep{jian14b, sado14, sado16, dai18, jian19}. These models explicitly predict `skin' outflows with velocities $\sim 0.1c$ \citep{dai18}. 

The high quality of spectropolarimetric data in SDSS~J1652 allows us to estimate the opening angle of the torus ($\sim 20-30\deg$) and the outflow velocity in the `skin' of the torus: in the data, the polarized peak is offset from the overall intensity peak by $450-850$ km s$^{-1}$, whereas in the models this offset is $\sim v_{\rm s}$. This angles agree well with the estimate of \citet{veil16} who constrain the line of sight in Mrk~231 to be within $\sim 10-26\deg$ of the polar axis. If the opening angle were much higher, then we cannot obtain an agreement between the position angle of polarization on the red wing of the line and that inferred from the orientation of the large-scale scattered light nebula \citep{vayn21a}. If the opening angle is smaller, then the geometry changes from equatorial to polar scattering with a relatively round projected scatterer and it becomes hard to produce a pronounced polarization angle swing or high levels of polarization. 

These estimates can now be compared from the results of theoretical models for near- and super-Eddington accretion -- although only in a qualitative sense, because the existing simulations tend to be focused on hyper-Eddington accretion. Nonetheless, the small opening angle of the torus, the low-density polar region and a higher density `skin' outflow clearly inferred from our data for SDSS~J1652 are in excellent qualitative agreement with theoretical expectations for the geometry of super-Eddington accretion. Our inferred outflow velocities are slower than those in simulations by about an order of magnitude likely because we are observing the outflow at distances of several tens of pc, as opposed to to the outflows seen in numerical simulations on scales of several tens of $R_g$, or $\sim 0.01$ pc. The outflow decelerates both due to its ballistic motion out of the black hole potential well and due to the entrainment of extra mass.

\subsection{Physical conditions in the scattering region}

We have presented electron scattering models as our primary models for comparison with observations. In Sec. \ref{sec:resonant}, we ruled out resonant scattering as the dominant scattering mechanism on the basis of the high polarization seen in the continuum and in emission lines which cannot resonantly scatter. This is somewhat surprising, given the high efficiency of the resonant scattering \citep{lee97, alex18} and the likely presence of the relevant ions in the scattering region. A possible explanation is geometric self-shielding: if the scattering medium is clumpy, with partially ionized clouds producing UV emission lines, then both line production and resonant scattering may be happening on the sides of the clouds facing the nucleus and not the observer. 

We further find in Sec. \ref{ssec:elec} that the high levels of polarization seen in SDSS~J1652 are in some tension with dust scattering and therefore electron scattering is preferred. Again, it is somewhat surprising: even in a fully ionized medium (as long as the conditions are not too harsh to destroy the dust), the efficiency of dust scattering is two orders of magnitude higher than the efficiency of Thomson scattering \citep{wein01}. It is possible that the polarization efficiency of the dust models can be increased somewhat by adjusting the size distribution, but any model with dominant forward-scattering will yield relatively low net polarization. The shape of the continuum of ERQs -- somewhat redder than that of type 1 quasars, but not consistent with single-layer dust reddening \citep{hama17} -- does not help in resolving dust vs electron scattering ambiguity, both because the incident continuum spectral shape in ERQs is not known and because the apparent wavelength dependence of scattering efficiency can be strongly affected by the number of scattering clouds and mixed-in reddening \citep{kish01}. 

We therefore discuss what kinds of physical conditions in the scattering outflow may make electron scattering possible. In observations of SDSS~J1652, the fractional polarization of emission lines maxes out at the same level ($\sim 20\%$) as the continuum polarization; this equality is achieved on the red wings of the lines. The net polarization of the blue wings of the lines is lower, likely due to the mixing of the continuum and line scattering with opposite Stokes values, as discussed in Sec. \ref{sec:analysis}. In contrast, in electron-scattering models the peak fractional polarization is significantly higher in the lines than in the continuum, reaching 40\% in the best geometric model in Figure \ref{pic:eq_within}. One possible real-world complication which would suppress the line polarization is the finite size of the emission region. In our models we assumed a point-like emission source which allows a major simplification of the calculations in that it is only necessary to integrate over the solid angles and not distances (Sec. \ref{sec:model_geometry}). In practice, as described in Sec. \ref{sec:dis_unif}, the outflowing skin of the torus is likely acting as both the emitter and the scatterer, so that the size of the scattering region is not that much larger than that of the emission region. Under these conditions, a well-known geometric cancellation lowers the average polarization level of the emission lines \citep{tran95b, zaka05}. 

The models of SDSS~J1652 in which the observer's line of sight is within the forward-scattering outflowing material naturally explain the net blueshift of the UV emission lines (Figure \ref{pic:1652}, top) compared to their wavelengths expected from the redshift determined by the optical forbidden emission lines in Gemini \citep{alex18} and {\it JWST} \citep{wyle22} data. The fact the polarized line flux is then redshifted by comparison with the total line flux is an indication that the scatterer is outflowing relative to the emitter. Thus our best-fitting models include only the hemisphere pointed toward the observer, which means we have implicitly assumed that the second hemisphere is highly obscured by circumnuclear dust. This is likely a safe assumption: even on much larger scales affected only by the obscuration of the host galaxy, the scattering and ionization counter-cone is much fainter than the cone directed toward the observer \citep{vayn21a, wyle22, vayn23b, vayn23c}. But in principle the presence of the backward-pointing hemisphere can be incorporated into the models by computing Stokes parameters for $\theta, \theta_{\rm s}<90\deg$ and $\theta_{\rm obs}>90\deg$ models and adding them to the corresponding $\theta_{\rm obs}<90\deg$ calculation. One can even include an adjustable extinction factor to mimic the partial obscuration of the back-facing flow. 

Based on their extremely high luminosity $L_{\rm bol}=10^{47}-10^{48}$ erg/sec, we suspect that extremely red quasars, of which SDSS~J1652 is an example, are near-Eddington objects. The necessary mass accretion rate to produce an Eddington luminosity at radiative efficiency $\varepsilon$ would be 
\begin{eqnarray}
\dot{M}_{\rm Edd}=\frac{4\pi GM_{\rm BH}m_p}{\varepsilon c \sigma_T}=\\
22\frac{M_{\odot}}{\rm year}\times\left(\frac{M_{\rm BH}}{10^9M_{\odot}}\right)\left(\frac{\varepsilon}{0.1}\right)^{-1}.
\label{eq_Edd}
\end{eqnarray}
In near-Eddington or super-Eddington sources we expect that the total kinetic energy of the outflow may constitute a significant fraction of the quasar's bolometric output and that that a non-negligible fraction of the mass fails to accrete and is ejected, with typical velocities $\sim 0.1c$ \citep{dai18}. Furthermore, the radiative efficiency of near-Eddington flows could be $\ll 0.1$. Therefore, eq. (\ref{eq_Edd}) likely provides a lower limit on the mass accretion rate. 

We now compare this accretion rate with the minimal mass ejection rate necessary to explain spectropolarimetric observations. We estimate the scattering efficiency -- the fraction of the photons originating near the nucleus that are then scatterer by the outflowing scattering material -- at $f_{\rm scat}=0.01$ based on the ratio of the UV flux observed in ERQs to that in an unobscured quasar of similar luminosity \citep{zaka16b}. This efficiency is proportional to the number density of particles and to the solid angle spanned by of the scatterer $\Delta \Omega$ (eq. 2 in \citealt{alex18}), but so is the outflowing mass $\dot{M}_{\rm out}=\Delta \Omega n_H m_p r_0^2 v_{\rm s}/X$, where $X=0.7$ is the hydrogen fraction by mass, $m_p$ is the proton mass and $r_0$ is the typical distance of the outflowing material from the source. Refurbishing this as a function of scattering efficiency, we find
\begin{equation}
\dot{M}_{\rm out}=\frac{8\pi m_p r_0 v_{\rm s} f_{\rm scat}}{(1+X)\sigma_{\rm T}g_{\rm TR}(\psi)}.
\end{equation}
Here we explicitly assumed that scattering is by electrons and are using Thomson scattering cross-section $\sigma_{\rm T}$. For dust scattering, the combination $\sigma_{\rm T}g_{\rm TR}(1+X)/(8\pi X)$ would need to be replaced by ${\rm d}C_{\rm scat}/{\rm d}\Omega(\psi)$, the differential cross-section for dust scattering per hydrogen atom at the dominant scattering angle $\psi$. 

In \citet{alex18}, we argued that $r_0$ should be around 10$-$30 pc based on the typical lengthscales of obscuration and dust sublimation. The velocity of the scatterer $v_{\rm s}$ is $\sim 800$ km s$^{-1}$ from comparing the results of our models with the observed offset between the peaks of the scattered and polarized emission; it is also just above the escape velocity from a $10^9M_{\odot}$ black hole at 30 pc. With these values at $\psi=30\deg$, we obtain
\begin{equation}
\dot{M}_{\rm out}=34\frac{M_{\odot}}{\rm year}\left(\frac{f_{\rm scat}}{0.01}\right)\left(\frac{r_0}{30{\rm pc}}\right)\left(\frac{v_{\rm s}}{800\mbox{km s}^{-1}}\right).
\label{eq:mass}
\end{equation}

This outflow rate is comparable to that expected near the black hole, where momentum conservation dictates $\dot{M}_{\rm wind}0.1c\simeq L/c$ \citep{king15} and therefore the outflowing mass is comparable to the Eddington rate. In practice the outflow at $10-30$ pc scales could be expected to be much more massive than near the nucleus due to entrainment of material and additional acceleration of winds by radiation pressure on the dust. We also expect these intermediate scale outflows we are probing with polarimetry to be clumpy, with the dense clouds dominating the mass budget but not necessarily dominating the scattering. The velocity dispersion $\sigma_v$ of scattered lines $\sim 1000$ km s$^{-1}$ constrains the temperature of the scattering electrons to be $\la m_e \sigma_v^2/(2 k_B)\simeq 3\times 10^4$ K \citep{zaka05}. Although this is unexpectedly cool for volume-filling, dust-free gas, there is precedent for such physical conditions: \citet{ogle03} demonstrated that the highly ionized gas with the direct view of the nucleus in NGC 1068 which serves as the scatterer for the optical emission has temperature $\la 10^5$ K. If the scattering is due to dust after all, then depending on the size distribution of particles, the required mass is at least an order of magnitude smaller than that given by eq. (\ref{eq:mass}). 

\section{Conclusions}
\label{sec:conc}

In this paper we present spectropolarimetric observations for an extremely red quasar SDSS~J1652 at $z=3$. The object is likely a near- or super-Eddington source, and it has a known powerful galaxy-wide quasar-driven ionized gas wind \citep{wyle22, vayn23b, vayn23c}. It is highly polarized in the rest-frame UV ($\sim 20\%$) and shows several intriguing features in the kinematic distribution of polarization within UV emission line. In particular, the polarization polarization angle changes dramatically within the lines, `swinging' from being aligned with the projected axis on the outflow on the blue side to a position angle that's perpendicular to the outflow axis on the red side and in the continuum. 

As near-Eddington sources in general and SDSS~J1652 in particular are known for their outflow activity on a variety of spatial scales, we develop a theoretical model of scattering which can explain the salient features of our observations with radial motions alone. We discuss key features of the observed polarized line profiles and their relationship with outflow geometry. Unsurprisingly, a polar outflow produces polarization with position angle perpendicular to the projected axis of the outflow, and this is not affected by the orientation of the observer. The velocity offset between the total scattered profile and the polarized profile can be used to probe the kinematics of the scatterer. In contrast, an equatorial outflow or a `torus skin' outflow can produce polarization either parallel or perpendicular to the projected axis, depending on the geometry of the outflow and orientation of the observer. 

The key spectropolarimetric properties of SDSS~J1652 -- the polarization position angle swing, the geometric relationship between the position angles of the UV polarization and the large-scale illuminated nebula and the kinematic offset between the polarized and the scattered line -- are well explained by a `skin outflow' model for the scatterer, with the observer's line of sight within the outflow. The typical inner opening angle of the outflow suggested by the models is $20-30\deg$, which is comparable to constraints for another super-Eddington source Mrk~231 based on other types of data \citep{veil16}. Skin winds from thick disks are in qualitative agreement with models of near- and super-Eddington accretion, although the observations probe them on much larger spatial scales (a few to a few tens of pc) than those that are accessible to numerical simulations (0.01 pc). 

The scattering mechanism remains difficult to pin down. Line resonant scattering does not appear to dominate as it cannot explain the high polarization of the continuum and of one particular line blend. Electron scattering is consistent with high observed values of polarization, but requires a relatively large scattering mass and dust-free scattering material at temperatures $\ll 10^5$K (otherwise the lines would be thermally broadened). Dust scattering requires only modest mass, but it is in tension with observed high values of polarization. The exact values depend on the particle size distribution, which is unknown, but any model of dust which results in strong forward scattering is likely to also result in low polarization, so it may be difficult to reach $\sim 20$\% polarization values in an outflow pointed directly toward the observer by adjusting the dust size distribution. Because most scattering results in polarization perpendicular to the scattering event plane, the qualitative kinematic features of the geometric model are independent of the exact scattering mechanism. 

\section*{Acknowledgements}

NLZ is grateful to R. Antonucci, B.T. Draine, J.E. Greene, J.F. Hennawi, J.H. Krolik, J.M. Stone and R.A. Sunyaev for useful discussions, and to the Institute for Advanced Study, Princeton, NJ for hospitality during the sabbatical when much of this work was done. The authors are grateful to S. Veilleux for permission to adapt his Mrk~231 cartoon and to the anonymous referee for the encouraging and constructive report. Support for this work was provided in part by the National Aeronautics and Space Administration (NASA) through Chandra Award Number GO6-17100X issued by the Chandra X-ray Observatory Center, which is operated by the Smithsonian Astrophysical Observatory for and on behalf of NASA under contract NAS8-03060. RMA was supported in part by NASA Jet Propulsion Laboratory subcontract 1520456 associated with the NASA Keck time allocation. NLZ was supported by the Catalyst Award of the Johns Hopkins University and by the Deborah Lunder and Alan Ezekowitz Founders' Circle Membership at the Institute for Advanced Study. 

The data presented herein were obtained at the W.M. Keck Observatory, which is operated as a scientific partnership among the California Institute of Technology, the University of California and NASA. The Observatory was made possible by the generous financial support of the W.M. Keck Foundation. The authors wish to recognize and acknowledge the very significant cultural role and reverence that the summit of Maunakea has always had within the indigenous Hawaiian community. We are most fortunate to have the opportunity to conduct observations from this mountain.

\section*{Data Availability}

The raw data for SDSS~J1652 used in this article are publicly available in the Keck archive. The numerical calculations and example models are available at \url{https://github.com/zakamska/polarized_outflows}. 

%\facility{Keck:I (LRIS)}

\bibliographystyle{mnras}
\bibliography{master}

\bsp	% typesetting comment
\label{lastpage}
\end{document}